\documentclass[showpacs,floats,floatfix,aps,prb,groupedaddress,showpacs,twocolumn,amsfonts,amssymb]{revtex4}

\usepackage{bm}
\usepackage{hyphenat}
\usepackage{graphicx}
\usepackage{psfrag}
\usepackage{psfig}
\usepackage[latin1]{inputenc}
\begin{document}
%\draft
\renewcommand{\thefootnote}{\fnsymbol{footnote}}

\title{Nature of Phase Transitions in a Generalized Complex $|\psi|^4$ Model}
\author{Elmar Bittner and Wolfhard Janke}

\affiliation{Institut f\"ur Theoretische Physik, Universit\"at Leipzig,
Augustusplatz 10/11, D-04109 Leipzig, Germany}

\begin{abstract}
\noindent 
We employ Monte Carlo simulations to study a generalized three-dimensional 
complex $|\psi|^4$ theory of Ginzburg-Landau form and compare our numerical
results with a recent 
quasi-analytical mean-field type approximation, which predicts first-order 
phase transitions in parts of the phase diagram. As we have shown earlier,
this approximation does not apply to the standard formulation of the model.
This motivated us to introduce a generalized Hamiltonian with an additional 
fugacity term controlling implicitly the vortex density. With this modification
we find that the complex $|\psi|^4$ theory can, in fact, be tuned to 
undergo strong first-order phase transitions. The standard model is confirmed 
to exhibit continuous transitions which can be characterized by XY model 
exponents, as expected by universality arguments. A few remarks on the 
two-dimensional case are also made.
\end{abstract}


\maketitle

\section{Introduction} \label{intro}

Since long the Ginzburg-Landau model has been considered as paradigm
for studying critical phenomena using field-theoretic 
techniques.\cite{ZinnJ}
Perturbative calculations of critical exponents and amplitude ratios
of the Ising ($n=1$), XY ($n=2$), Heisenberg ($n=3$) and other O($n$) spin 
models relied heavily on this field-theoretic formulation.\cite{KlSch01}
Even though the spin models contain only directional fluctuations, while for
$n$-component Ginzburg-Landau fields with $n\ge2$ directional and size
fluctuations seem to be equally important, the two descriptions are completely
equivalent, as is expected through the concept of universality and has been 
proved explicitly for superfluids with $n=2$,
where the spin model reduces to an XY model.\cite{Kl00}
Therefore it appeared as a surprise when,
on the basis of an approximate variational approach to the two-component
Ginzburg-Landau model, Curty and Beck \cite{beck1} recently predicted for
certain parameter ranges the possibility of first-order phase transitions
induced by phase fluctuations. In several papers
\cite{beck2,fort2d_0,fort2da,fort2db,fort3d} this
quasi-analytical \cite{footnote1}
prediction was tested by Monte Carlo simulations and, as the main result,
apparently confirmed numerically. If true, these findings would have
an enormous impact on the theoretical description of many related systems
such as superfluid helium, superconductors, certain liquid crystals and
possibly even the electroweak standard model of elementary particle
physics.\cite{Kl89,LoQuSh01}
 
In view of these potential important implications for a broad variety of
different fields we performed independent Monte Carlo simulations of the
standard Ginzburg-Landau 
model in two and three dimensions in order to test whether
the claim of phase-fluctuation induced first-order transitions is a real
effect or not.\cite{ebwj_prl} Our results clearly support the prevailing 
opinion that the
nature of the transition is of second order. In turn this implies,
of course, that the variational approximation employed in 
Ref.~\onlinecite{beck1} 
is less reliable than originally thought in view of the apparent numerical
confirmations. 
In order to shed some light on the numerical results of 
Refs.~\onlinecite{beck2,fort2d_0,fort2da,fort2db,fort3d}, we generalized the
standard model by adding a fugacity term which implicitly controls the
vortex density of the model. The purpose of this paper is to present
for this generalized Ginzburg-Landau model results on its phase structure as
obtained from extensive Monte Carlo simulations. Employing finite-size scaling
analyses we find numerical evidence that, by tuning the extra fugacity 
parameter, it is indeed possible to drive the system into a region with 
first-order phase transitions. 

The layout of the remainder of this paper is organized as follows. In 
Sec.~\ref{model} we
first recall the standard model, and then discuss its generalization and the 
observables used to map out the phase diagram. 
Next we describe the employed simulation techniques in Sec.~\ref{numer}.
The results of our simulations are presented in Sec.~\ref{results}, where we
first discuss the three-dimensional case in some detail and then add a few 
brief comments on the two-dimensional model to complete the physical picture.
Finally, in Sec.~\ref{summary} we conclude with a summary of our main findings.

%%%%%%%%%%%%%%%%%%%%%%%%%%%%%%%%%%%%%%%%%%%%%%%%%%%%%%%%%%%%%%%%%%%%%%%%%%
\section{Model and Observables} \label{model} 
%%%%%%%%%%%%%%%%%%%%%%%%%%%%%%%%%%%%%%%%%%%%%%%%%%%%%%%%%%%%%%%%%%%%%%%%%%

The standard complex or two-component Ginzburg-Landau theory is defined by the
Hamiltonian
\begin{equation}
H[\psi] = \int \!\! \mathrm{d}^dr \left[\alpha |\psi|^2 + \frac{b}{2}|\psi|^4 +
\frac{\gamma}{2}|\nabla \psi|^2 \right], \quad \gamma > 0~,
\label{eq:H}
\end{equation}
where $\psi(\vec{r}) = \psi_x(\vec{r}) + i  \psi_y(\vec{r}) = 
|\psi(\vec{r})| e^{i \phi(\vec{r})}$ is a complex
field, and $\alpha$, $b$ and $\gamma$ are temperature independent coefficients 
derived from a microscopic model.
In order to carry out Monte Carlo simulations we put
the model (\ref{eq:H}) on a $d$-dimensional hypercubic lattice with spacing $a$.
Adopting the notation of Ref.~\onlinecite{beck1}, we introduce scaled variables
$\tilde{\psi} = \psi/\sqrt{(|\alpha|/b)}$ and $\vec{u}=\vec{r}/ \xi$,
where $\xi=\sqrt{\gamma/|\alpha|}$ is the mean-field correlation length at zero
temperature. This leads to the normalized lattice Hamiltonian
\begin{equation}\label{h2}
H[\tilde{\psi}] = k_B \tilde{V}_0  \sum_{n=1}^N \Big [\frac{\tilde{\sigma}}{2}
(|\tilde{\psi}_n|^2 - 1)^2 +
\frac{1}{2}\sum_{\mu=1}^d |\tilde{\psi}_n-\tilde{\psi}_{n+\mu}|^2 \Big ]~,
\end{equation}
with 
\begin{equation}
\tilde{V}_0=\frac{1}{k_B}\frac{|\alpha|}{b}\gamma a^{d-2}~,\quad
\tilde{\sigma}=\frac{a^2}{\xi^2}~,
\end{equation}
where $\mu$ denotes the unit vectors along the $d$ coordinate axes, 
$N=L^d$ is the total
number of sites, and an unimportant constant term has been removed. The
parameter $\tilde{V}_0$ merely sets the temperature scale and can thus be
absorbed in the definition of the reduced temperature 
$\tilde{T} = T/\tilde{V}_0$.

After these rescalings, and omitting the tilde on $\psi$, $\sigma$, and $T$ 
for notational simplicity in the rest of the paper,
the partition function $Z$ considered in the simulations is then given by
\begin{equation}
Z=\int \!\! D\psi D\bar{\psi} \, e^{-H/T}~,
\label{eq:Z}
\end{equation}
where
\begin{equation}
H[\psi] = \sum_{n=1}^N \Big [\frac{\sigma}{2}
(|\psi_n|^2 - 1)^2 +
\frac{1}{2}\sum_{\mu=1}^d |\psi_n-\psi_{n+\mu}|^2 \Big ]
\label{eq:H_scal}
\end{equation}
and $\int D\psi \,D\bar{\psi} \equiv \int D\,{\rm Re\/}\psi \,D\,{\rm Im\/}\psi$
stands short for integrating over all possible complex field configurations.

In Ref.~\onlinecite{ebwj_prl} we have shown, that the disagreement mentioned above
is caused by an incorrect sampling of the Jacobian which emerges from the complex 
measure in (\ref{eq:Z}) when transforming the field representation
to polar coordinates, $\psi_n = R_n (\cos(\phi_n), \sin(\phi_n))$.
When updating in the simulations the modulus $R_n = |\psi_n|$ and the
angle $\phi_n$, one has to rewrite the measure of the partition function 
(\ref{eq:Z}) as
\begin{equation}
Z=\int_0^{2\pi} \!\! D\phi \int_0^\infty \!\! R DR \, e^{-H/T}~,
\label{eq:Z_R}
\end{equation}
where $DR \equiv \prod_{n=1}^N dR_n$ and $R \equiv \prod_{n=1}^N R_n$ is the Jacobian 
of this transformation. While mathematically indeed trivial (and of course properly
taken into account in Ref.~\onlinecite{beck1}), this fact may
easily be overlooked when coding the update proposals for the modulus and angle
in a Monte Carlo simulation program. While for the angles it is correct to
use update proposals of the form $\phi_n \rightarrow \phi_n + \delta \phi$ with
$-\Delta \phi \le \delta \phi \le \Delta \phi$ (where $\Delta \phi$ is chosen such 
as to assure an optimal acceptance ratio), the similar procedure for the modulus,
$R_n \rightarrow R_n + \delta R$ with $-\Delta R\le \delta R \le \Delta R$,
would be incorrect since this ignores the $R_n$ factor coming from the Jacobian.
In fact, if we purposely ignore the Jacobian and simulate the model (\ref{eq:Z_R})
(erroneously) without the $R$-factor,
then we obtain a completely different behavior than in the correct case, 
cf.\ e.g.\ Fig.~\ref{pr} below. As already mentioned
above these results reproduce \cite{footnote2} those in
Refs.~\onlinecite{beck2} and \onlinecite{fort3d}, and from this data one 
would indeed
conclude evidence for a first-order phase transition when $\sigma$ is small.
With the correct measure, on the other hand, we have checked that {\em no}
first-order signal shows up down to $\sigma = 0.01$.

To treat the measure in Eq.~(\ref{eq:Z_R}) properly one can either use the
identity $R_n dR_n = d R_n^2/2$ and update the squared moduli 
$R_n^2 = |\psi_n|^2$ according to a uniform measure (where the update proposal
$R_n^2 \rightarrow R_n^2 + \delta$ with $-\Delta \le \delta \le \Delta$ is 
correct), or one can introduce an effective Hamiltonian,
\begin{equation}
H_{\rm eff} = H - T \kappa \sum_{n=1}^N \ln R_n~,
\label{eq:Heff}
\end{equation}
with $\kappa \equiv 1$ and work directly with a uniform measure for $R_n$.
The incorrect omission of the
$R$-factor in (\ref{eq:Z_R}) is equivalent to setting $\kappa = 0$. It is well
known \cite{Kl89} that the nodes $R_n=0$ correspond to core regions of
vortices in the dual formulation of the model. The Jacobian factor $R$
(or equivalently the term $-\sum \ln R_n$ in $H_{\rm eff}$) tends
to suppress field configurations with many nodes $R_n=0$. If the $R$-factor
is omitted, the number of nodes and hence vortices is relatively
enhanced. It is thus at least qualitatively plausible that in this case
a discontinuous, first-order ``freezing transition'' from a vortex dominated
phase can occur, as is suggested by a similar mechanism for the 
XY model \cite{JK1,Kl89,MiWa87} and defect-models of 
melting \cite{Kl89II,WJ_defect}.

In the limit of a large parameter $\sigma$, it is
easy to read off from Eq.~(\ref{eq:H_scal}) that the modulus of the
field is squeezed onto unity and once hence expects that irrespectively of 
the value of $\kappa$ the XY model limit is approached with its well-known 
continuous phase transition in three dimensions (3D) at $T_c \approx 2.2$ 
respectively Kosterlitz-Thouless (KT) transition in two dimensions (2D) 
at $T_{\rm KT} \approx 0.9$. 
While for the standard model with $\kappa = 1$, this 
behavior should qualitatively persist for all values of $\sigma$, 
from the numerical results discussed above one expects that for $\kappa = 0$
the order of the transition turns first-order below a certain (tricritical) 
$\sigma$-value. The purpose of this paper is to elucidate this 
behavior further by studying the phase diagram in the $\sigma$-$\kappa$-plane,
i.e., by considering an interpolating model with $\kappa$ varying 
continuously between 0 and 1. 

To be precise we always worked with the proper functional measure in 
Eq.~(\ref{eq:Z_R}) and replaced the standard Hamiltonian $H$ by
\begin{equation}
H_{\rm gen} = H + T(1-\kappa) \sum_{n=1}^N \ln R_n
            = H + T \delta \sum_{n=1}^N \ln |\psi_n|~,
\label{eq:H_gen}
\end{equation}
where we have introduced the parameter $\delta = 1 - \kappa$,
such that $\delta = 0$ ($\kappa = 1$) corresponds to the standard model and
$\delta = 1$ ($\kappa = 0$) to the previously studied modified model with its
first-order phase transition for small enough $\sigma$.

In order to map out the phase diagram in the $\sigma$-$\kappa$-
respectively $\sigma$-$\delta$-plane, 
we have measured 
in our simulations to be described in detail in the next section
among other quantities the energy density $e=\langle H\rangle / N$,
the specific heat per site $c_v=(\langle H^2\rangle- \langle H\rangle^2)/N$, and
in particular the mean-square amplitude
\begin{equation}
\langle|\psi|^2\rangle=\frac{1}{N} \sum_{n=1}^N
\langle |\psi_n|^2\rangle~,
\label{eq:mean_square}
\end{equation}
which will serve as the most relevant quantity for comparison with previous
work \cite{beck1,beck2,fort2d_0,fort2da,fort2db,fort3d}. For further
comparison and in order to determine the critical temperature, the helicity
modulus,
\begin{eqnarray}
\Gamma_\mu = \frac{1}{N}\langle \sum_{n=1}^N
|\psi_n||\psi_{n+\mu}|
\cos(\phi_n - \phi_{n+\mu})\rangle \nonumber\\
-\frac{1}{NT} \langle \left[\sum_{n=1}^N|\psi_n|
|\psi_{n+\mu}| \sin(\phi_n - \phi_{n+\mu}) \right]^2 \rangle~,
\label{eq:helicity}
\end{eqnarray}
was also computed.
Notice that the helicity modulus $\Gamma_\mu$ is a direct measure of the
phase correlations in the direction of $\mu$. Because of cubic symmetry all
directions $\mu$ are equivalent, and we always quote the average
$\Gamma = (1/d) \sum_{\mu=1}^d \Gamma_\mu$. 
In the infinite-volume limit,
$\Gamma$ is zero above $T_c$ and different from zero below $T_c$.
We also have measured the vortex density $v$ (of vortex points in 2D and 
vortex lines in 3D). The standard procedure to calculate
the vorticity on each plaquette is by considering the quantity 
\begin{equation}
m=\frac{1}{2\pi}([\phi_1-\phi_2]_{2\pi}+[\phi_2-\phi_3]_{2\pi}+[\phi_3-\phi_4]_{2\pi}+[\phi_4-\phi_1]_{2\pi})~,
\end{equation}
where $\phi_1,\dots,\phi_4$ are the phases at the corners of a plaquette labeled,
say, according to the right-hand rule, and
$[\alpha]_{2\pi}$ stands for $\alpha$ modulo $2\pi$: 
$[\alpha]_{2\pi}=\alpha+2\pi n$,
with $n$ an integer such that $\alpha+2\pi n \in (-\pi,\pi]$, hence 
$m=n_{12}+n_{23}+n_{34}+n_{41}$. If $m\neq0$, there exists a vortex 
which is assigned to the object dual to the given plaquette (a site in 2D and a link in 3D).
Hence, in two dimensions,
${*m}$, the dual of $m$, is assigned to the center of the original plaquette. 
In three dimensions, the topological point charges are replaced by
(oriented) line elements ${*l_i}$ which combine to form closed networks 
(``vortex loops''). The vortex ``charges'' ${*m}$ or ${*l_i}$ can take three 
values: $0,\pm 1$ (the values $\pm 2$ have a 
negligible probability). The quantities
\begin{eqnarray} 
v &=& \frac{1}{L^2}\sum_{x}|{*m}_x| \quad {\rm (2D)}~,
\label{eq:vortex2D} \\
v &=& \frac{1}{L^3}\sum_{x,i}|{*l}_{i,x}| \quad {\rm (3D)}
\label{eq:vortex3D}
\end{eqnarray}
serve as a measure of the vortex density.
We further analyzed the Binder cumulant,
\begin{equation}
U=\frac{\langle (\vec{\mu}^2)^2 \rangle}{\langle\vec{\mu}^2\rangle^2}~,
\end{equation}
where $\vec{\mu}=(\mu_x, \mu_y)$ with
\begin{equation}
\mu_x=\frac{1}{N}\sum_{n=1}^{N} {\rm Re}(\psi_n)~,\quad 
\mu_y=\frac{1}{N}\sum_{n=1}^{N} {\rm Im}(\psi_n)~,
\end{equation}
is the magnetization per lattice site of a given configuration.

%%%%%%%%%%%%%%%%%%%%%%%%%%%%%%%%%%%%%%%%%%%%%%%%%%%%%%%%%%%%%%%%%%%%%%%%%%
\section{Simulation Techniques} \label{numer} 
%%%%%%%%%%%%%%%%%%%%%%%%%%%%%%%%%%%%%%%%%%%%%%%%%%%%%%%%%%%%%%%%%%%%%%%%%%

Let us now turn to the description of the Monte Carlo update procedures used
by us. To be on safe grounds,
we started with the most straightforward (but most inefficient)
algorithm known since the early days of Monte Carlo simulations: The standard
Metropolis algorithm \cite{Metro}. Here the complex field $\psi_n$ is
decomposed into its
Cartesian components, $\psi_n = \psi_{x,n} + i \psi_{y,n}$. For each lattice
site a random update proposal for the two components is made, e.g.
$\psi_{x,n} \rightarrow \psi_{x,n} + \delta \psi_{x,n}$ with
$\delta \psi_{x,n} \in [-\Delta,\Delta]$, and in the standard fashion accepted
or rejected according to the energy change $\delta H_{\rm gen}$. The parameter $\Delta$
is usually chosen such as to give an acceptance rate of about $50\%$, but
other choices are permissible and may even result in a better performance of
the algorithm (in terms of autocorrelation times). All this is
standard \cite{WJ_review} and guarantees in a straightforward manner that the
complex measure $D\psi D\bar{\psi}$ in the partition function (\ref{eq:Z})
is treated properly.

\begin{figure}[t]
\centerline{\psfig{figure=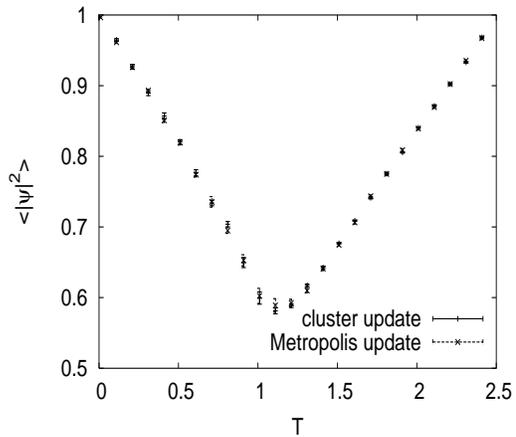,angle=0,height=6.cm,width=7cm}}
\caption{\label{fig:metro}
Mean-square amplitude of the standard three-dimensional complex 
Ginzburg-Landau model with 
$\kappa = 1$ and $\sigma = 0.25$ on a $10^3$ cubic lattice.
}
\end{figure}

The well-known drawback of this algorithm is its critical slowing down
(large autocorrelation times) in the vicinity of a continuous phase
transition \cite{WJ_review}, leading to large statistical errors for a fixed
computer budget. To improve the accuracy of our data we therefore employed
the single-cluster algorithm \cite{wolff} to update the direction of the
field \cite{hasen}, similar to simulations of the XY spin model \cite{WJ_XY}.
The modulus of $\psi$
is updated again with a Metropolis algorithm. Here some care is necessary
to treat the measure in (\ref{eq:Z}) properly (see above comments).
Per measurement we performed one sweep with the Metropolis algorithm and $n$
single-cluster updates. For all simulations in two and three dimensions
the number of cluster updates was chosen such that $n \langle |C| \rangle
\approx L^d \equiv N$, where $\langle |C| \rangle$ is the average cluster size. 
Since $\langle |C| \rangle$ scales with system size 
as the susceptibility, $\chi = N \langle \vec{\mu}^2 \rangle \simeq 
L^{\gamma/\nu}$, and 
$\gamma/\nu = 2 - \eta = 7/4$ at the Kosterlitz-Thouless transition in 2D 
and $\gamma/\nu = 2 - \eta \approx 2$ in
3D, $n$ was chosen $\propto L^{1/4}$ in 2D and $\propto L$ in 3D.
In the 2D case most of the simulations were performed for $L=10, 20$, and $40$, 
and in 3D we usually studied the lattice sizes $L=10, 15, 20$, and $30$. For 
each simulation point we thermalized with 500 to $1\,000$ sweeps and averaged 
the measurements over $10\,000$ sweeps.
In the cases of strong first-order phase transitions we employed
a variant of the multicanonical scheme~\cite{berg} where the
histogram of the mean modulus is flattened instead that of the energy.
All error bars are computed with the Jackknife method \cite{Jack}. In the
following we
only show the more extensive and accurate data set of the cluster simulations,
but we tested in many representative cases that the Metropolis simulations
coincide within error bars, for an example see Fig.~\ref{fig:metro}.

\begin{figure*}[htb]
\centerline{\hbox{\psfig{figure=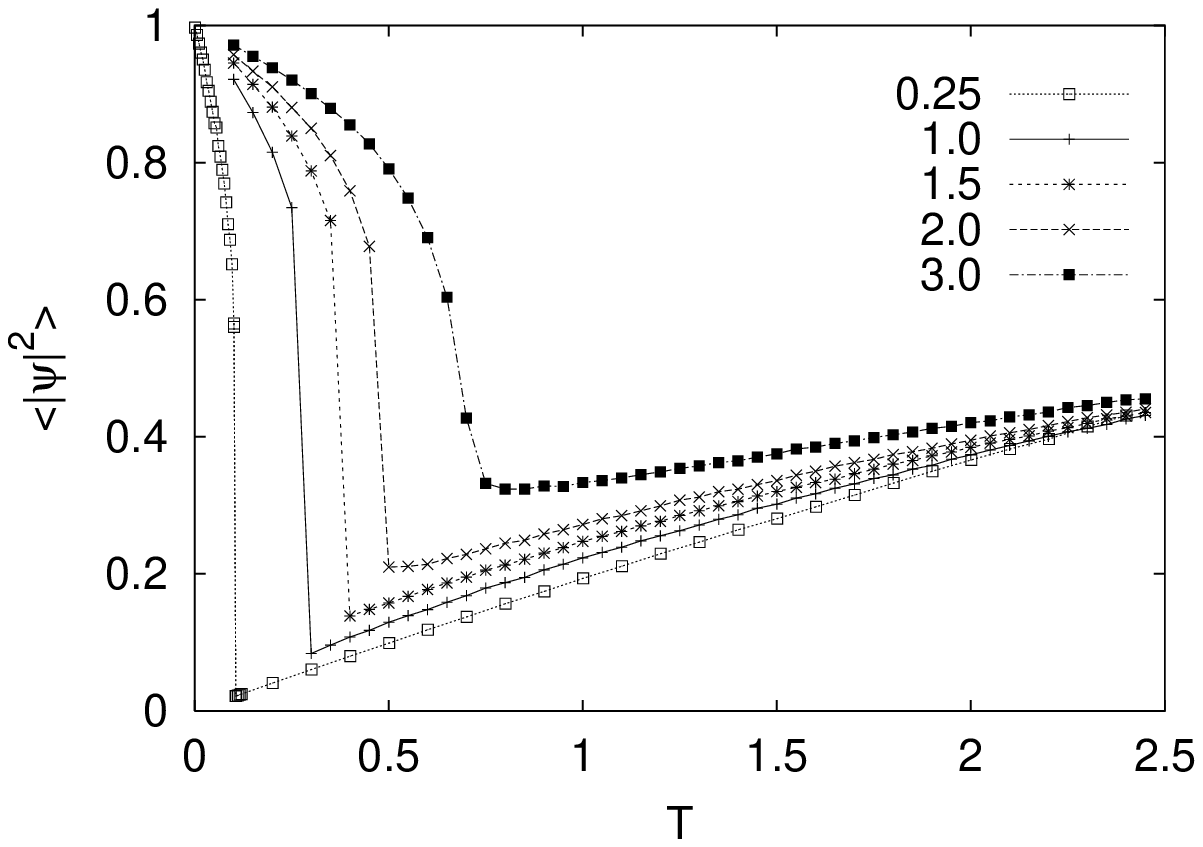,angle=0,height=6.cm,width=7cm}
\hspace*{1cm}
\psfig{figure=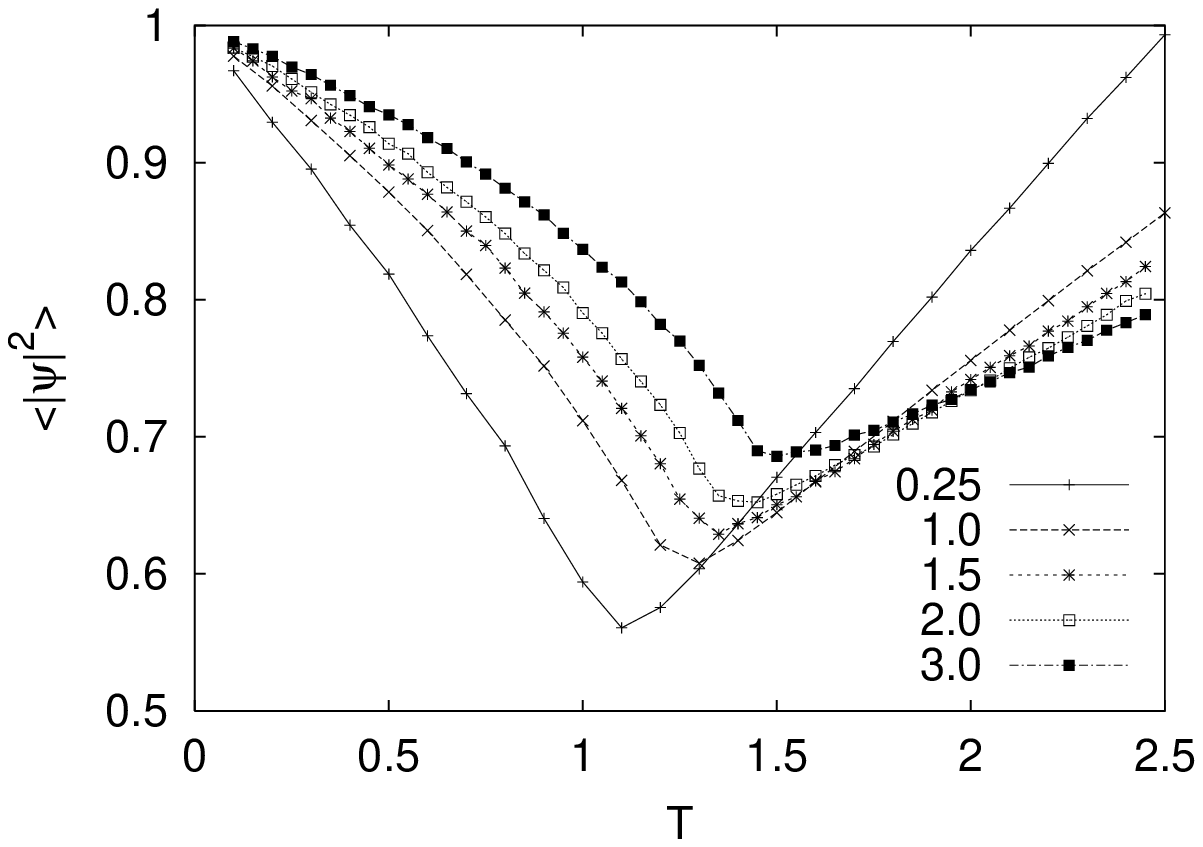,angle=0,height=6.cm,width=7cm}}}
\centerline{\hbox{\psfig{figure=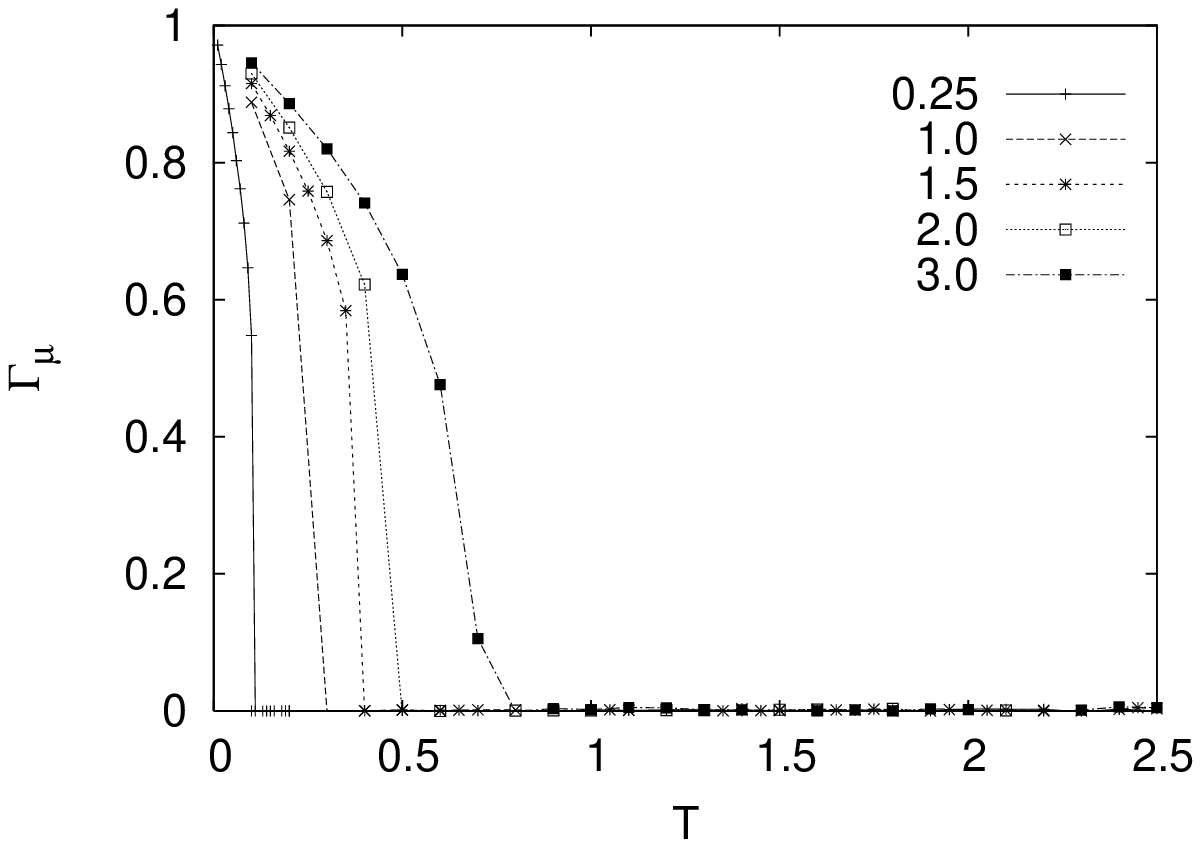,angle=0,height=6.cm,width=7cm}
\hspace*{1cm}
\psfig{figure=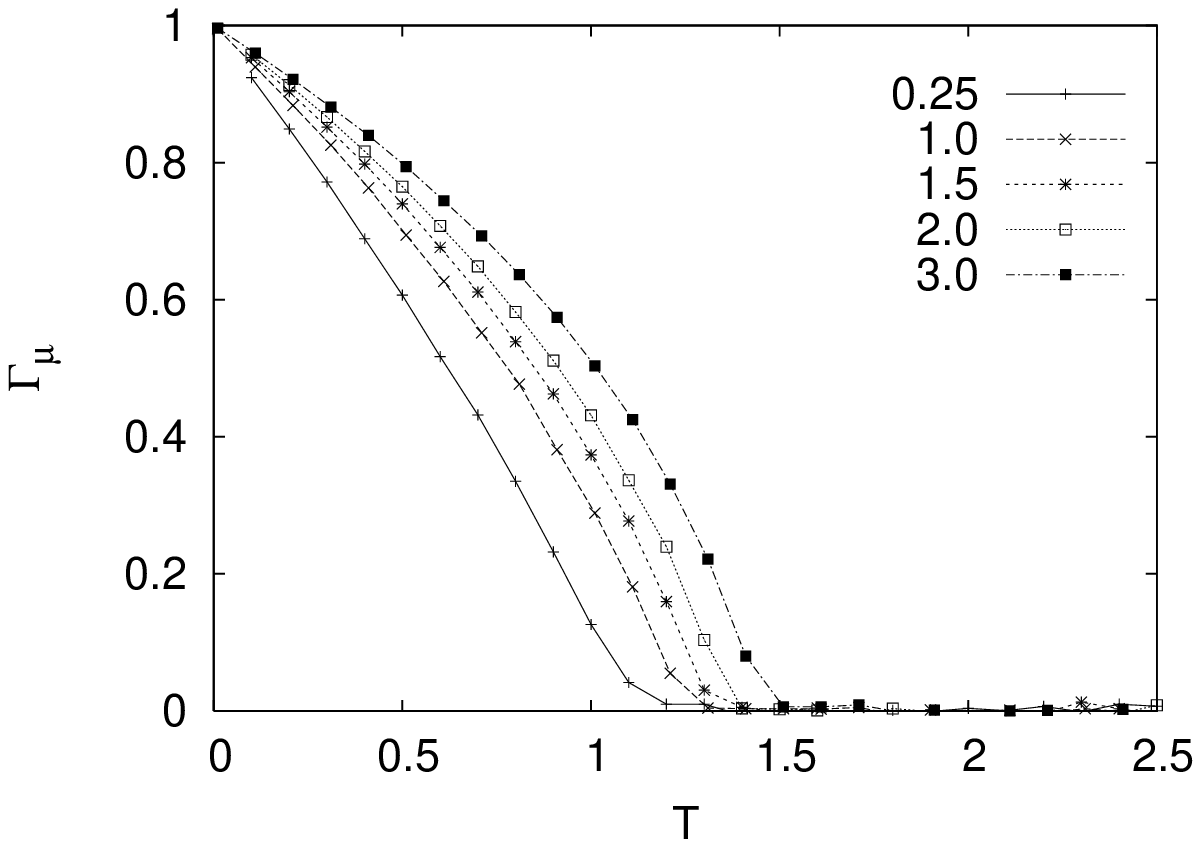,angle=0,height=6.cm,width=7cm}}}
\centerline{\hbox{\psfig{figure=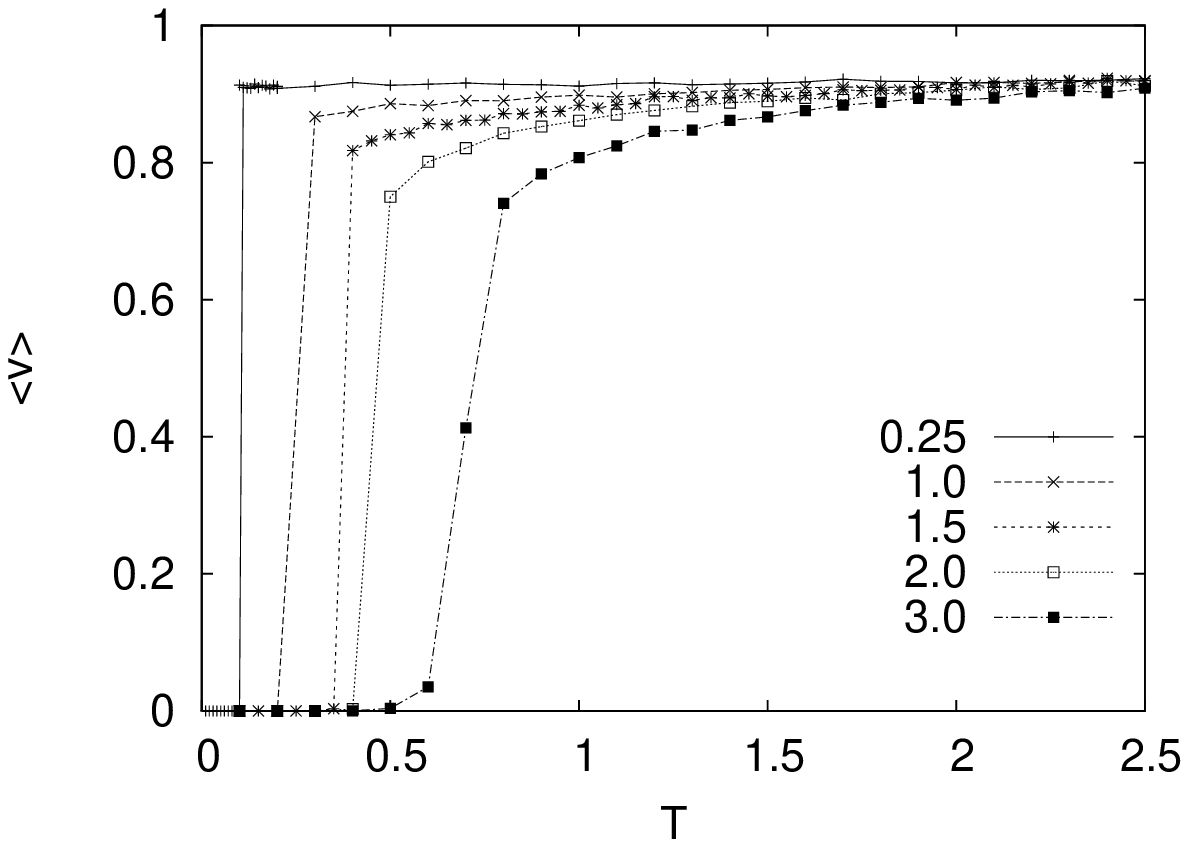,angle=0,height=6.cm,width=7cm}
\hspace*{1cm}
\psfig{figure=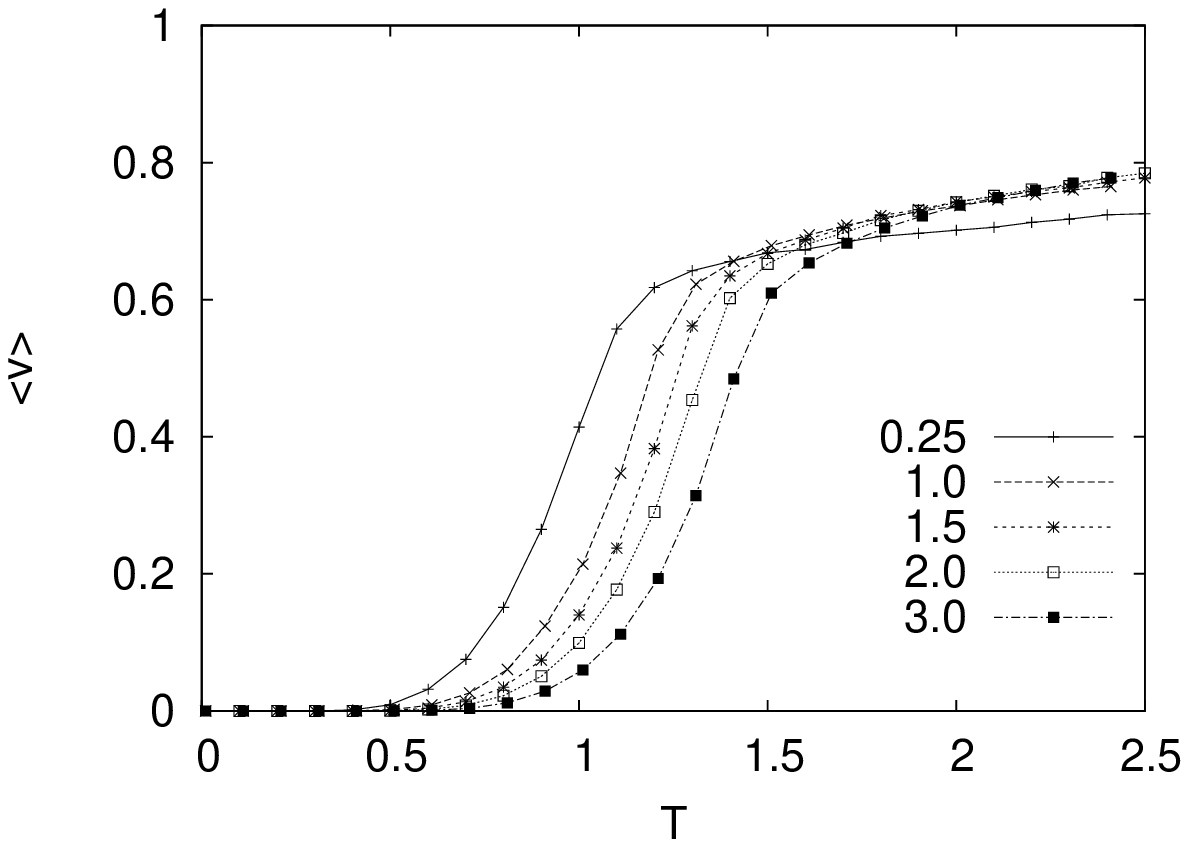,angle=0,height=6.cm,width=7cm}}}
\caption{\label{pr}
Mean-square amplitude $\langle |\psi|^2 \rangle$, helicity modulus 
$\Gamma_{\mu}$
and vortex-line density $\langle v \rangle$ 
of the  three-dimensional generalized complex Ginzburg-Landau model on 
a $15^3$ cubic lattice 
for different values of the parameter $\sigma = 0.25, \dots, 3.0$ for the
case $\kappa=0$ (left) and the standard formulation with $\kappa=1$ (right).
}
\end{figure*}
%

%%%%%%%%%%%%%%%%%%%%%%%%%%%%%%%%%%%%%%%%%%%%%%%%%%%%%%%%%%%%%%%%%%%%%%%%%%
\section{Results} \label{results} 
%%%%%%%%%%%%%%%%%%%%%%%%%%%%%%%%%%%%%%%%%%%%%%%%%%%%%%%%%%%%%%%%%%%%%%%%%%

\subsection{Three dimensions}

\begin{figure}[t]
\centerline{\hbox{\psfig{figure=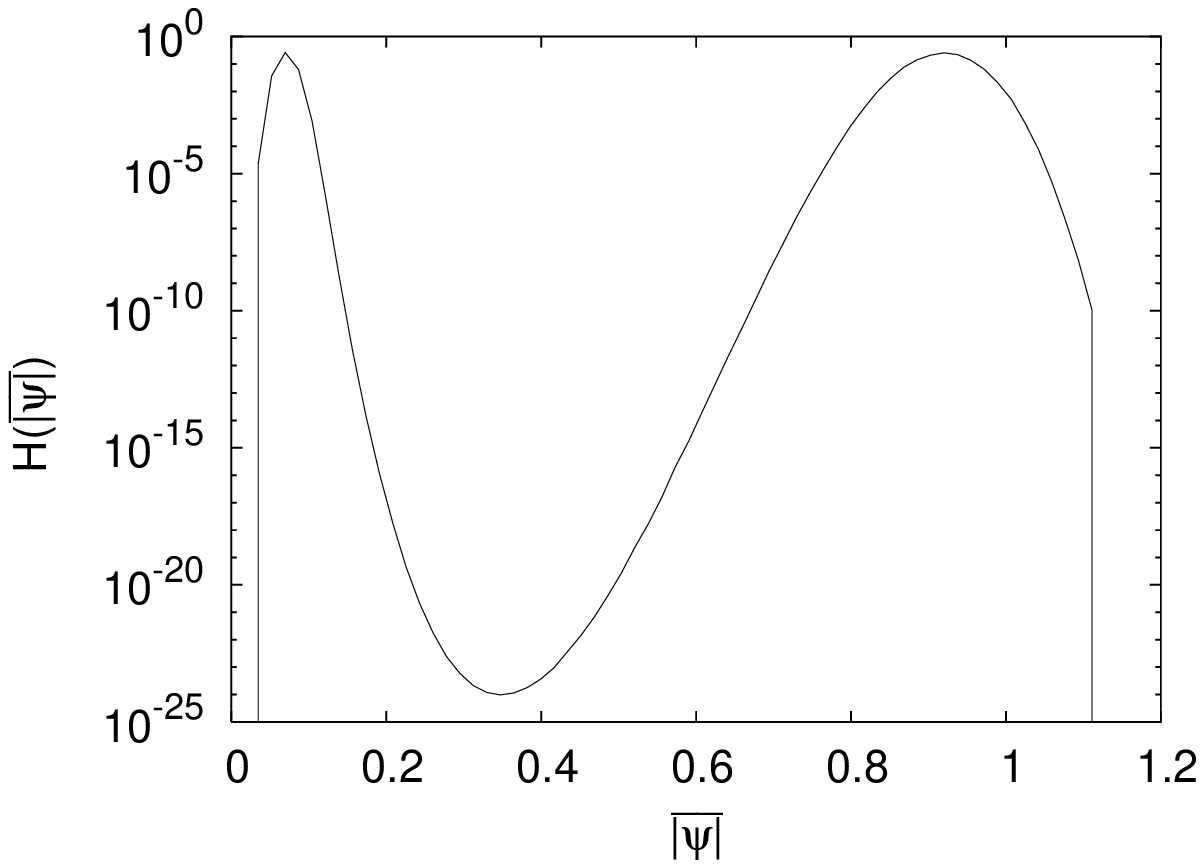,angle=0,height=6.cm,width=7cm}}}
\centerline{\hbox{\psfig{figure=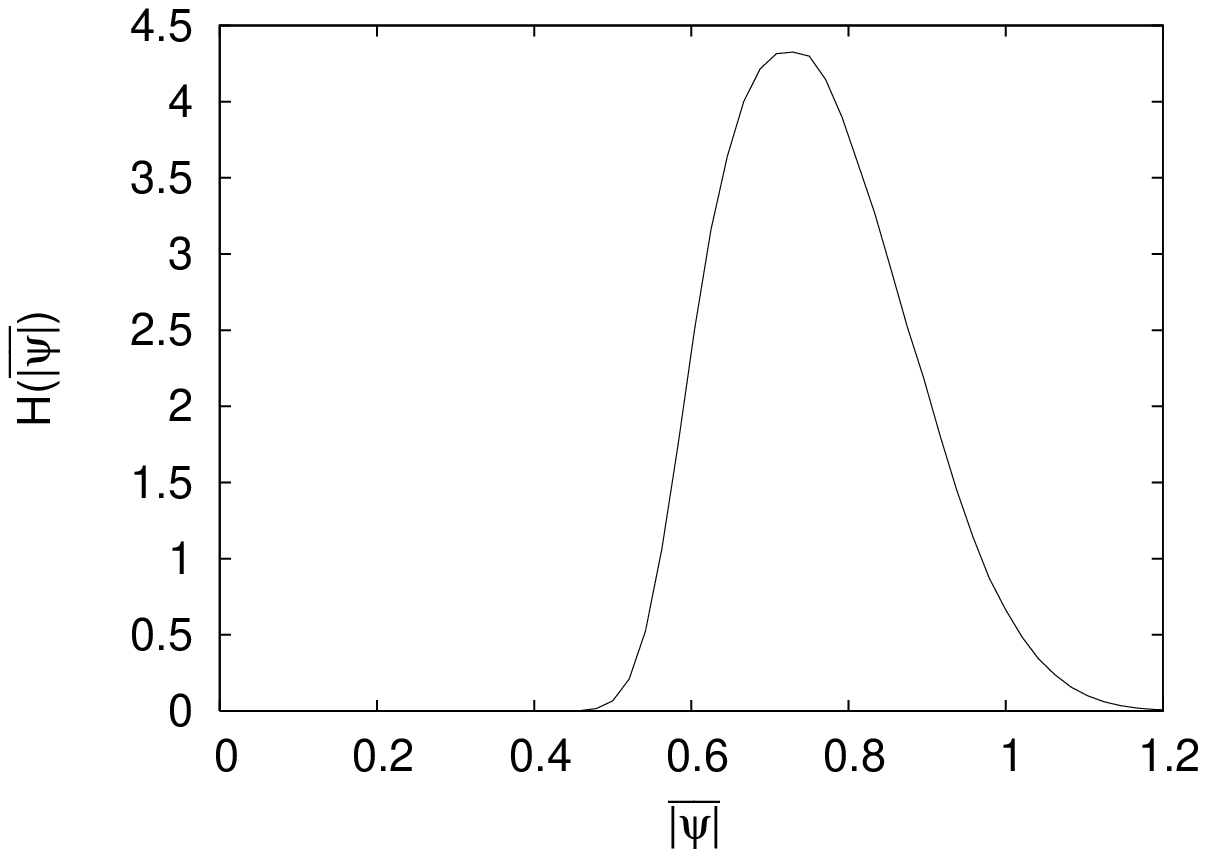,angle=0,height=6.cm,width=7cm}}}
\caption{\label{fig:histos}
Top: Histogram of the mean modulus $\overline{|\psi|}$ on a
logarithmic scale for a $4^3$ cubic lattice, $\kappa=0$ and $\sigma=0.25$, 
reweighted to the temperature $T_0 \approx 0.0572$ where the two peaks are 
of equal height.
Bottom: Histogram for the same quantity and lattice size at $T=1.1$ 
close to the second-order phase transition
for $\kappa=1$ and $\sigma=0.25$.
}
\end{figure}

\begin{figure}[t]
\centerline{\hbox{\psfig{figure=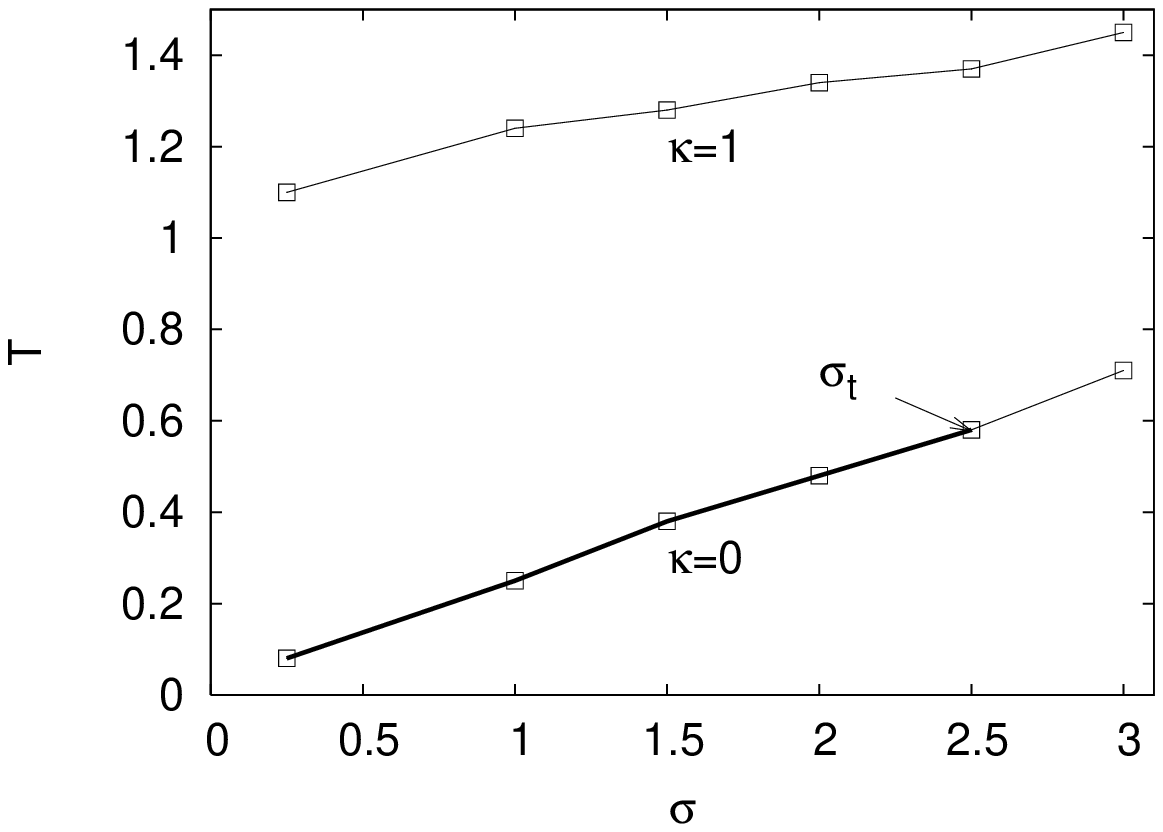,angle=0,height=6.0cm,width=7.0cm}}}
\caption{\label{fig:sigma_T_diag}
Transition lines in the $\sigma$-$T$-plane for $\kappa=0$ and $\kappa=1$.
The thick line for $\kappa = 0$ indicates first-order phase transitions
while all other transitions are continuous.
}
\end{figure}

In the first set of simulations we concentrated on the two most
characteristic cases $\kappa = 0$ and $\kappa = 1$ and performed 
temperature scans on a $15^3$ lattice for various values of the
parameter $\sigma$. Our results for the mean-square amplitude, the
helicity modulus and the vortex-line density are compared for the two cases
in Fig.~\ref{pr}. In the plots for $\kappa = 0$ on the left side, we see that
all three quantities exhibit quite pronounced jumps for small
$\sigma$-values, which is a clear indication that in this regime the 
phase transition is of first order. At $\sigma = 0.25$, for example, we
observe already on very small lattices a clear double-peak structure
for the distributions of the energy and mean-square amplitude as well as
the mean modulus $\overline{|\psi|} = \frac{1}{N} \sum_{n=1}^N |\psi_n|$ 
which is depicted in Fig.~\ref{fig:histos}. Notice that already for the
extremely small lattice size of $4^3$ the minimum between the two peaks 
is suppressed by more than 20 orders of magnitude. This is an 
unambiguous indication for two coexisting phases and thus clearly implies that 
the model undergoes a first-order phase transition in the small $\sigma$-regime
for $\kappa=0$. 
Due to the pronounced metastability these
simulations had to
be performed with a variant of the multicanonical scheme~\cite{berg} where,
instead of flattening the energy histogram, extra weight factors for the
mean modulus were introduced. With this simulation technique we overcome the
difficulty of sampling the extremely rare events between the two peaks of the
canonical distribution. 
A closer look at the $\kappa=0$ plots shows that the
crossover from second- to first-order transitions happens around 
$\sigma_t \approx 2.5$. 
For the standard model with $\kappa = 1$, on the other hand, we 
observe for {\em all\/} $\sigma$-values a smooth behavior, suggesting that 
the XY model like continuous transition persists also for small $\sigma$-values.
This is clearly supported by a single-peak structure of all
distributions just mentioned, for the case of the mean modulus see 
Fig.~\ref{fig:histos}. This supports the prevailing opinion that the 
standard complex $|\psi|^4$ model always undergoes a second-order phase 
transition.
In fact, we have checked that down to $\sigma = 0.01$ {\em no\/} signal of a 
first-order transition can be detected for the standard model parameterized
by $\kappa = 1$. 
The resulting transition lines in the $\sigma$-$T$-plane for $\kappa=0$ and
$\kappa = 1$ are sketched in Fig.~\ref{fig:sigma_T_diag}, with the thick line
for $\kappa = 0$ indicating the approximate regime of first-order phase
transitions.

\begin{figure*}[tbh]
\centerline{\hbox{\psfig{figure=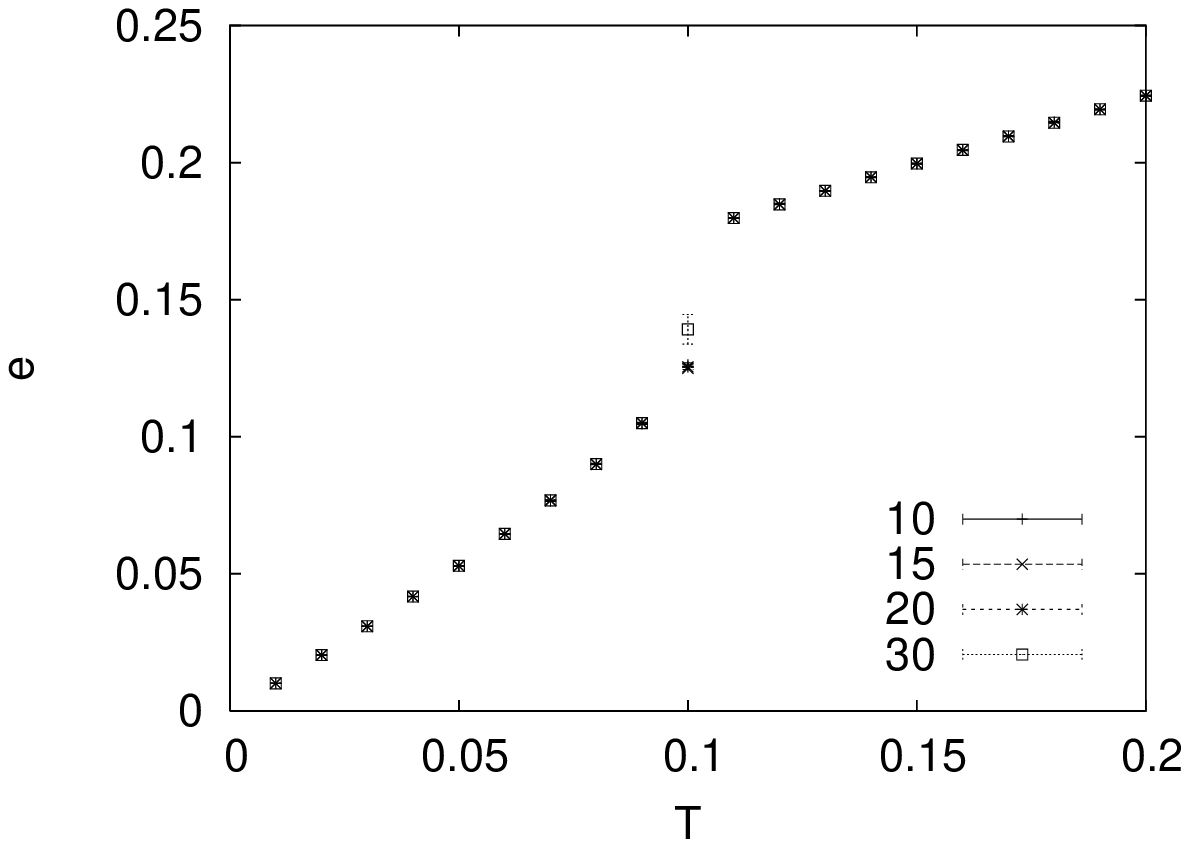,angle=0,height=6.0cm,width=7.0cm}
                  \psfig{figure=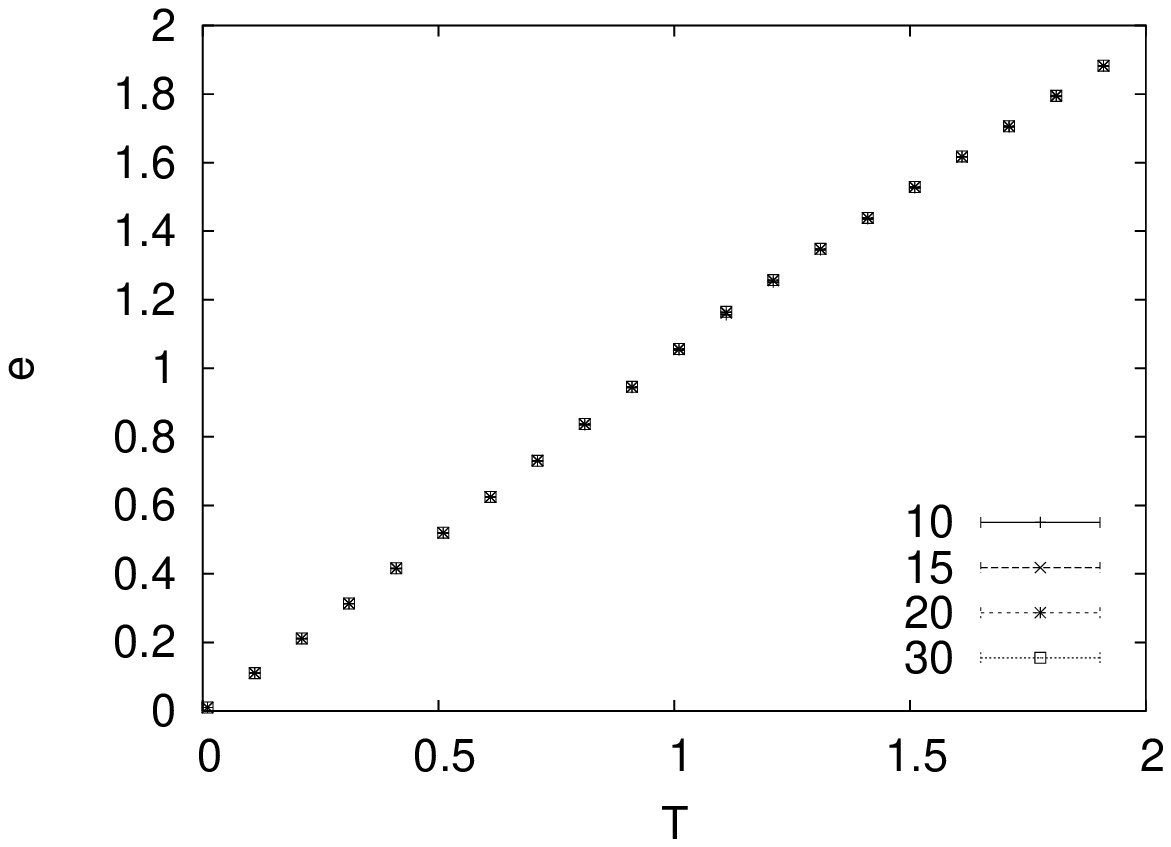,angle=0,height=6.0cm,width=7.0cm}}}
\centerline{\hbox{\psfig{figure=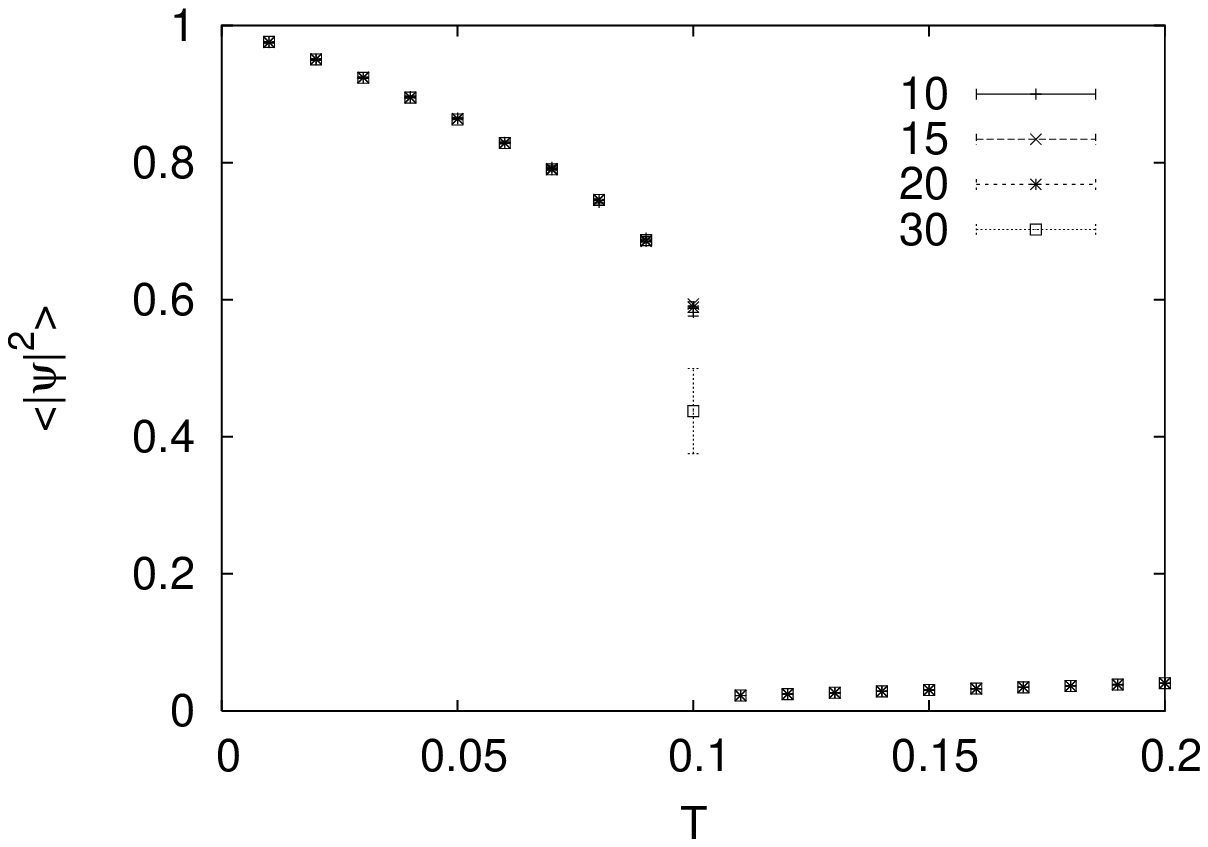,angle=0,height=6.0cm,width=7.0cm}
                  \psfig{figure=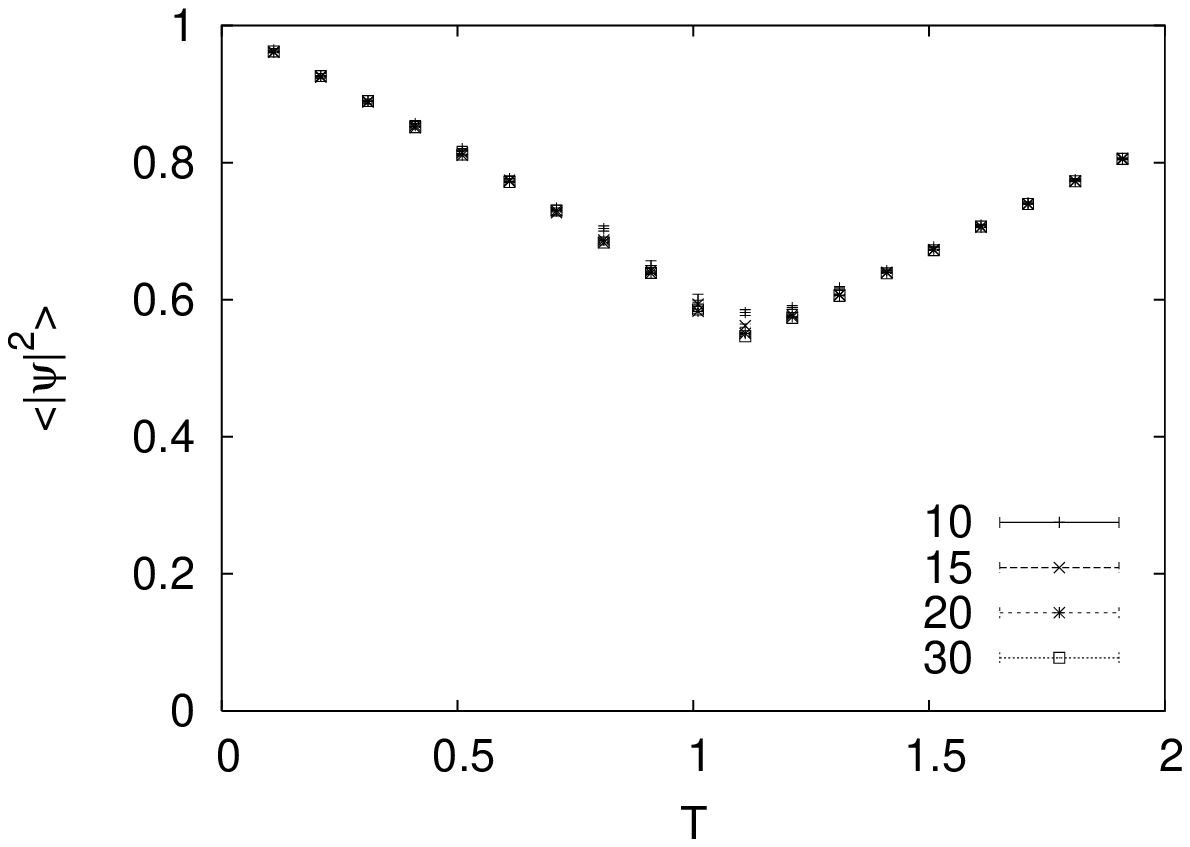,angle=0,height=6.0cm,width=7.0cm}}}
\centerline{\hbox{\psfig{figure=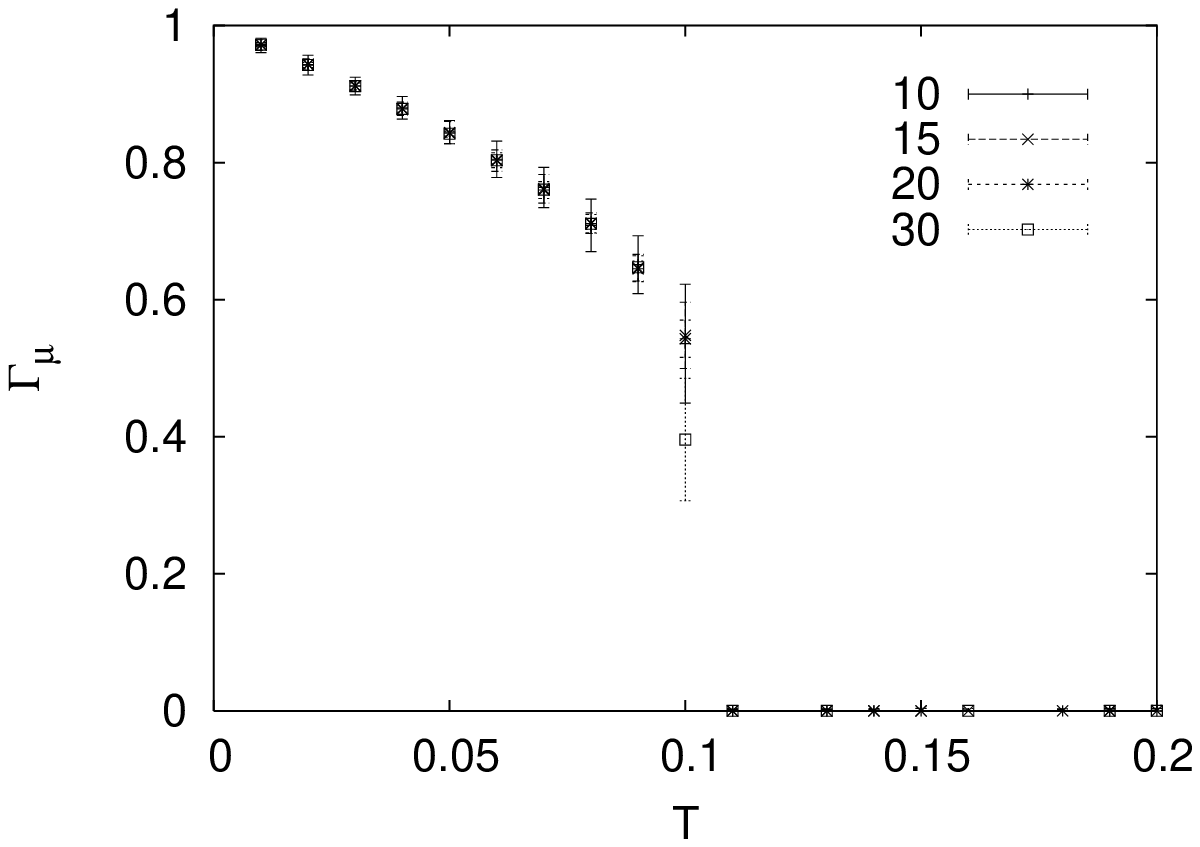,angle=0,height=6.0cm,width=7.0cm}
                  \psfig{figure=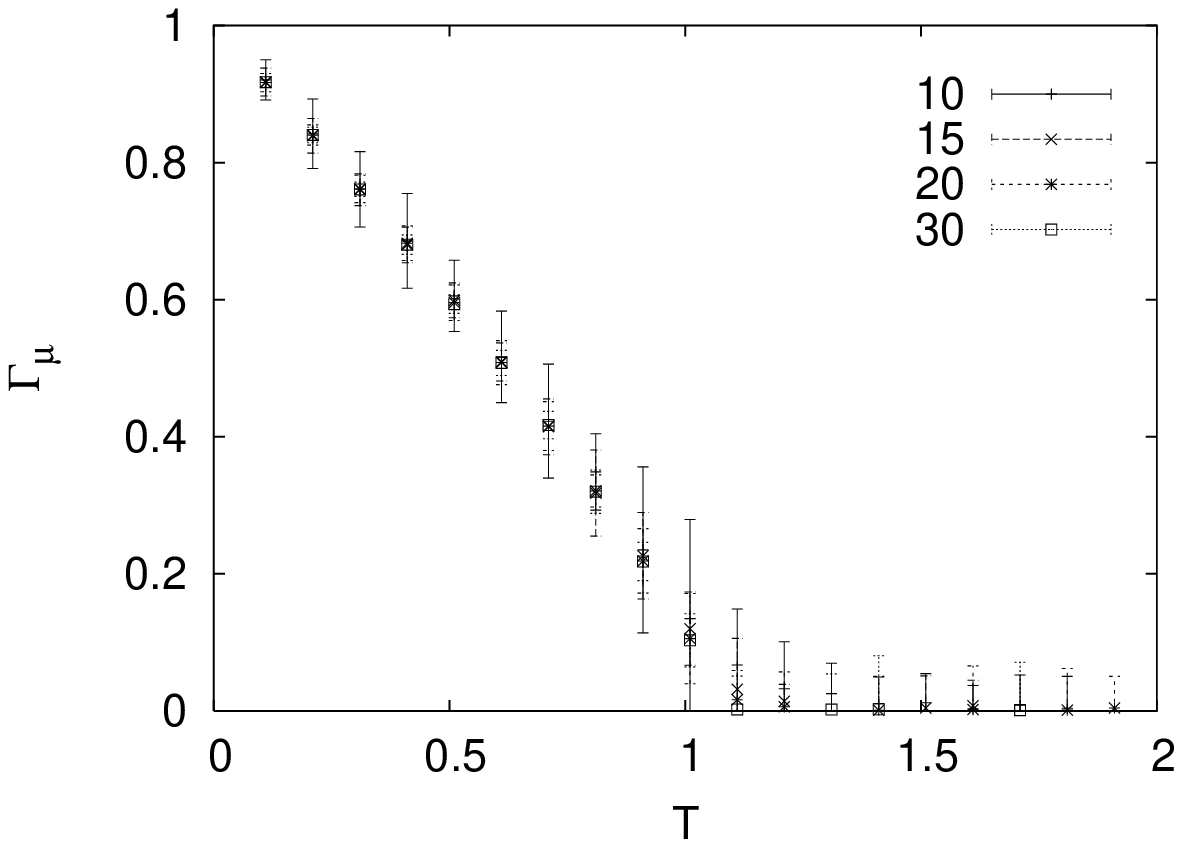,angle=0,height=6.0cm,width=7.0cm}}}
\centerline{\hbox{\psfig{figure=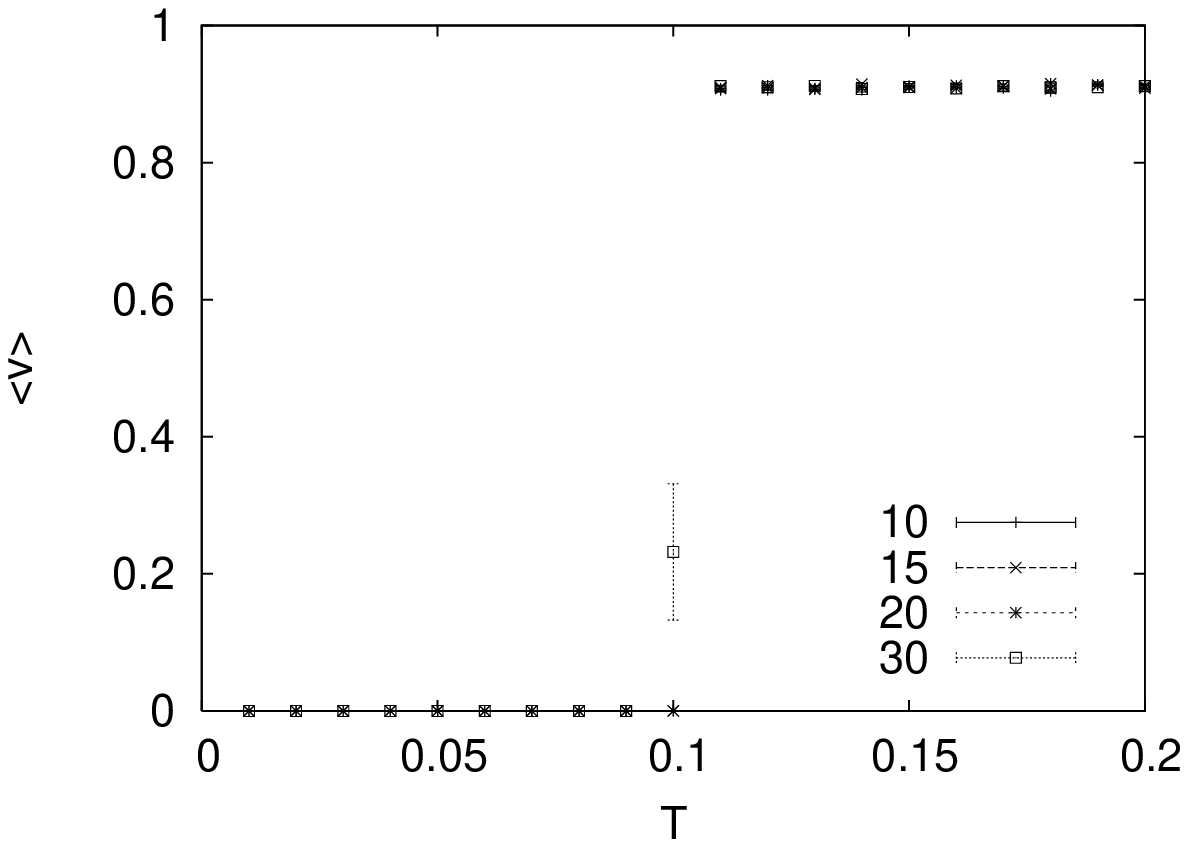,angle=0,height=6.0cm,width=7.0cm}
                  \psfig{figure=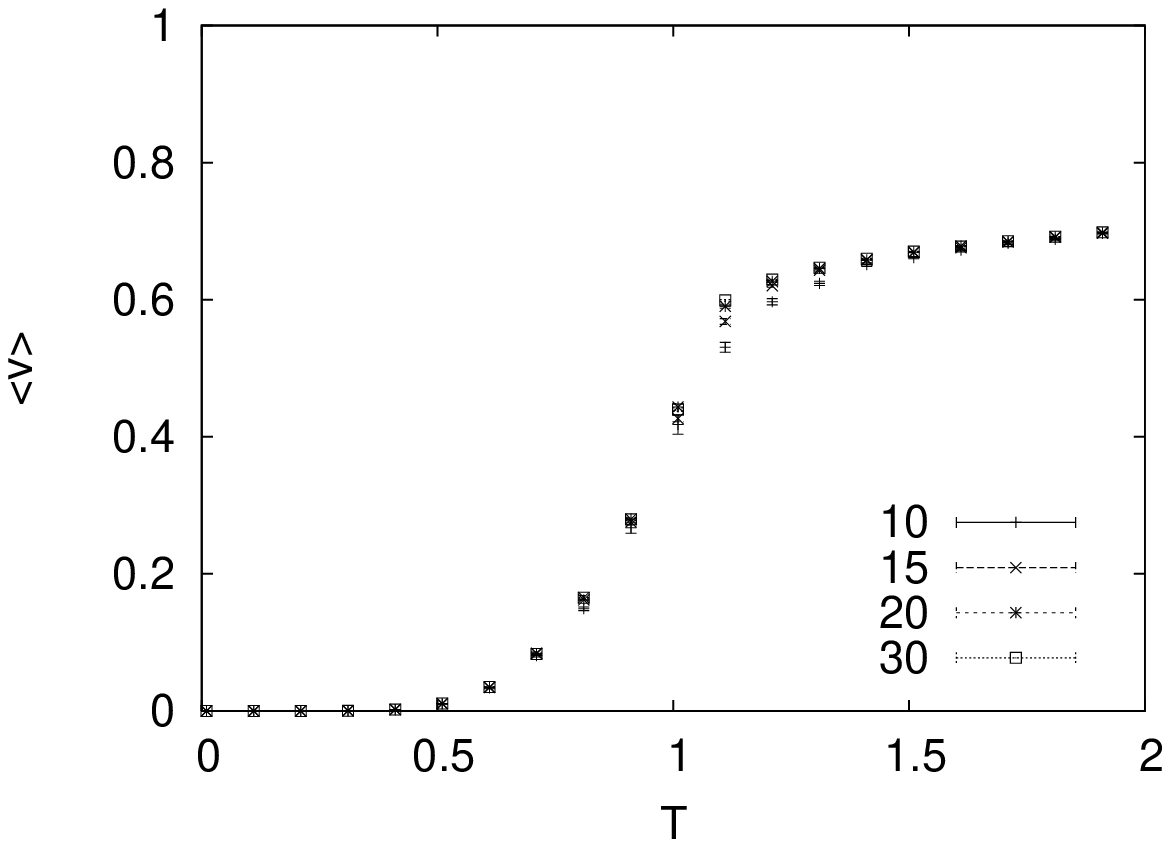,angle=0,height=6.0cm,width=7.0cm}}}
\caption{\label{fig_e}
Energy density $e$, mean-square amplitude $\langle|\psi|^2\rangle$,
helicity modulus $\Gamma_\mu$ and vortex-line density $v$
on $10^3, 15^3, 20^3$ and $30^3$ cubic lattices for $\sigma = 0.25$ and 
$\kappa = 0$ (left) respectively $\kappa = 1$ (right).
}
\end{figure*}

Next we concentrated on the small $\sigma$ regime and performed a rough
finite-size scaling (FSS) analysis for $\sigma = 0.25$ on moderately large 
$10^3$, $15^3$, $20^3$, and $30^3$ lattices.
In Fig.~\ref{fig_e} we compare results for the energy, mean-square
amplitude (\ref{eq:mean_square}), helicity modulus (\ref{eq:helicity}) 
and vortex-line density (\ref{eq:vortex3D}) for $\kappa=0$ and $\kappa=1$. 
Apart from the transition region where a strong size dependence is of course
expected, we notice only a small dependence on the variation of the 
lattice size. On the basis of these results, we do not expect a significant 
change of the qualitative behavior for much larger lattices and hence used 
similar moderate lattice sizes for most of our further investigations.

\begin{figure}[t]
\centerline{\psfig{figure=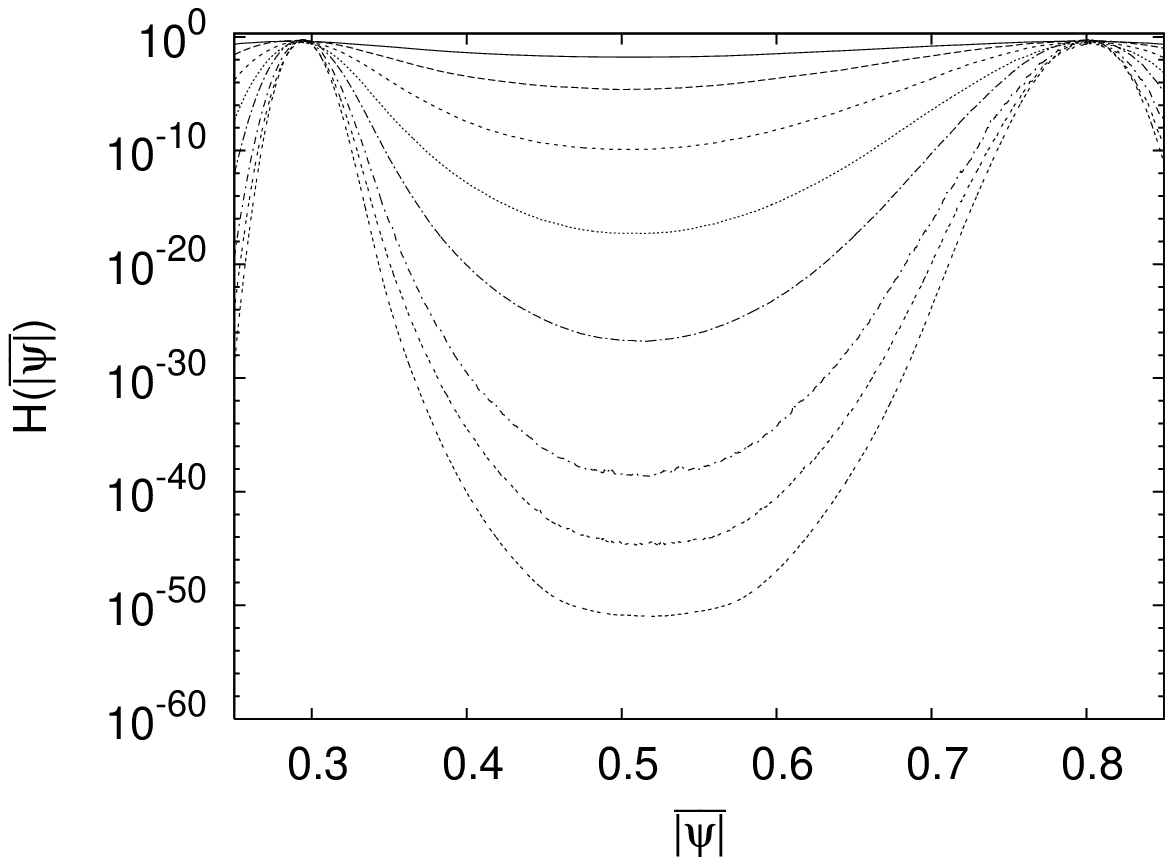,angle=0,height=6.cm,width=7cm}}
\centerline{\psfig{figure=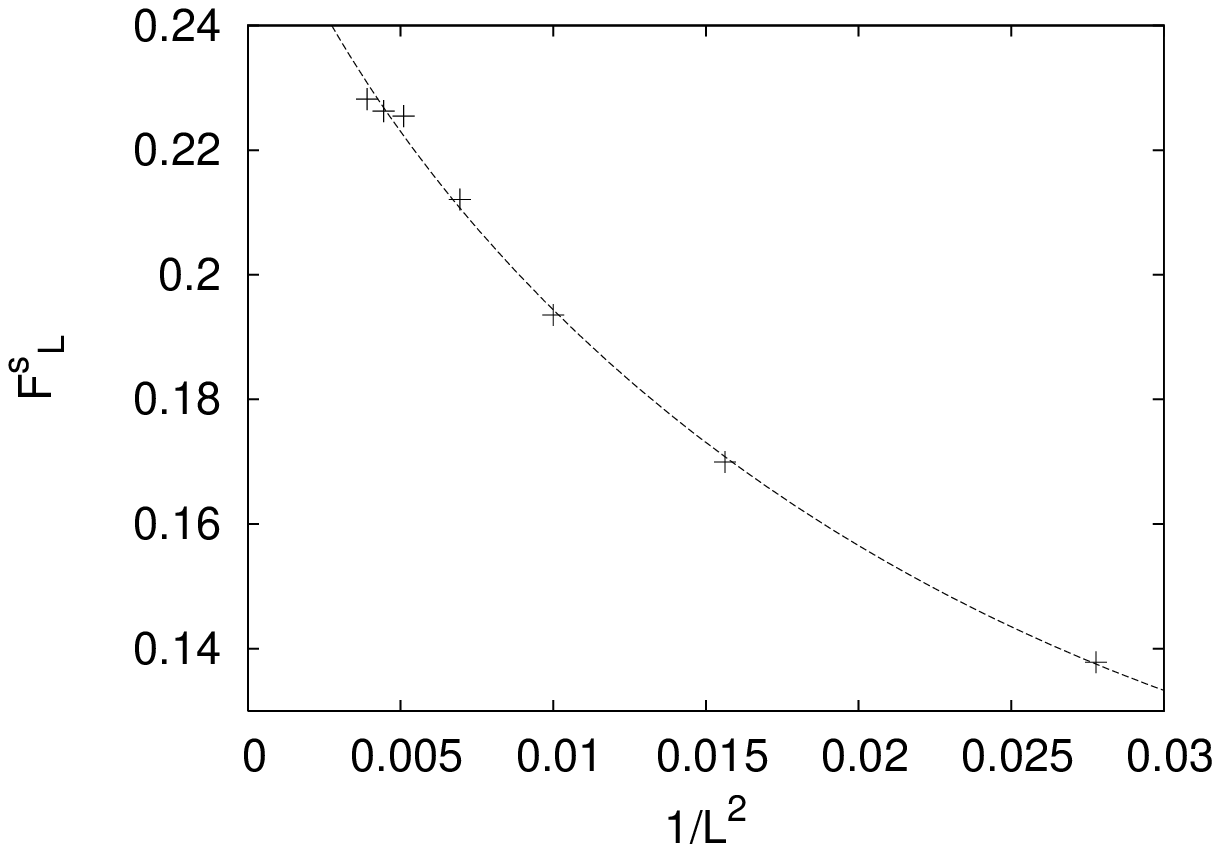,angle=0,height=6.cm,width=7cm}}
\caption{\label{fig:inter}
Top: Histogram of the mean modulus
$\overline{|\psi|}$ for $\kappa = 0$ and $\sigma = 1.5$ on a
logarithmic scale for various lattice sizes ranging from $L=4$ (top curve) to
$L=16$ (bottom curve), reweighted to temperatures where the two peaks
are of equal height.
Bottom: FSS extrapolation for $L \ge 6$ of the interface tension $F^s_L$, 
yielding the infinite-volume limit $F^s = 0.271(5)$.
}
\end{figure}

To exemplify the big differences between the models with $\kappa=0$ and 
$\kappa=1$, we choose in the following the case $\sigma = 1.5$,
where we shall characterize for both $\kappa$-values the phase transitions 
in some detail. Let us start with the non-standard case $\kappa=0$, 
where the first-order phase transition around $T \approx 0.36$ is also 
pronounced but much less strong than for $\sigma=0.25$. Still, in order to 
get sufficiently accurate equilibrium results, the
simulations for lattices of size $L=4,6,8,10,12,14,15$, and $16$ had to
be performed again with our modulus variant of the multicanonical method. 
As can be inspected in the histogram plots for the mean modulus
shown in Fig.~\ref{fig:inter}, the frequency of the rare events between the two peaks
in the canonical ensemble for a 
$16^3$ lattice is about 50 orders of magnitude smaller than for configurations
contributing to the two peaks.

\begin{figure}[htb]
\centerline{\hbox{\psfig{figure=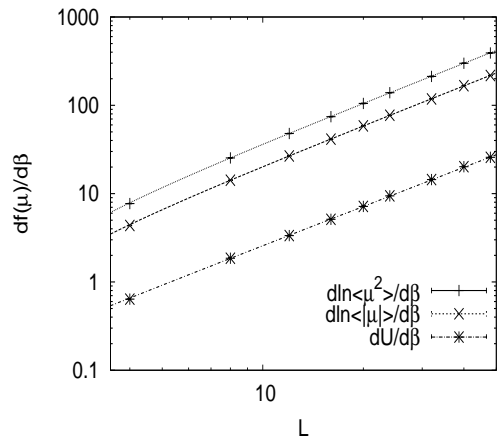,angle=0,height=6.cm,width=7cm}}}
\caption{\label{fig_nu}
Least-square fits for $\kappa = 1$ and $\sigma = 1.5$ on a log-log scale,
using the FSS ansatz
$d f(\mu)/d\beta \propto L^{1/\nu}$ at the maxima locations.
The fits using the data for $L \ge 8$ lead
to an overall critical exponent $1/\nu=1.493(7)$ or $\nu = 0.670(3)$.
}
\end{figure}

\begin{figure}[htb]
\centerline{\hbox{\psfig{figure=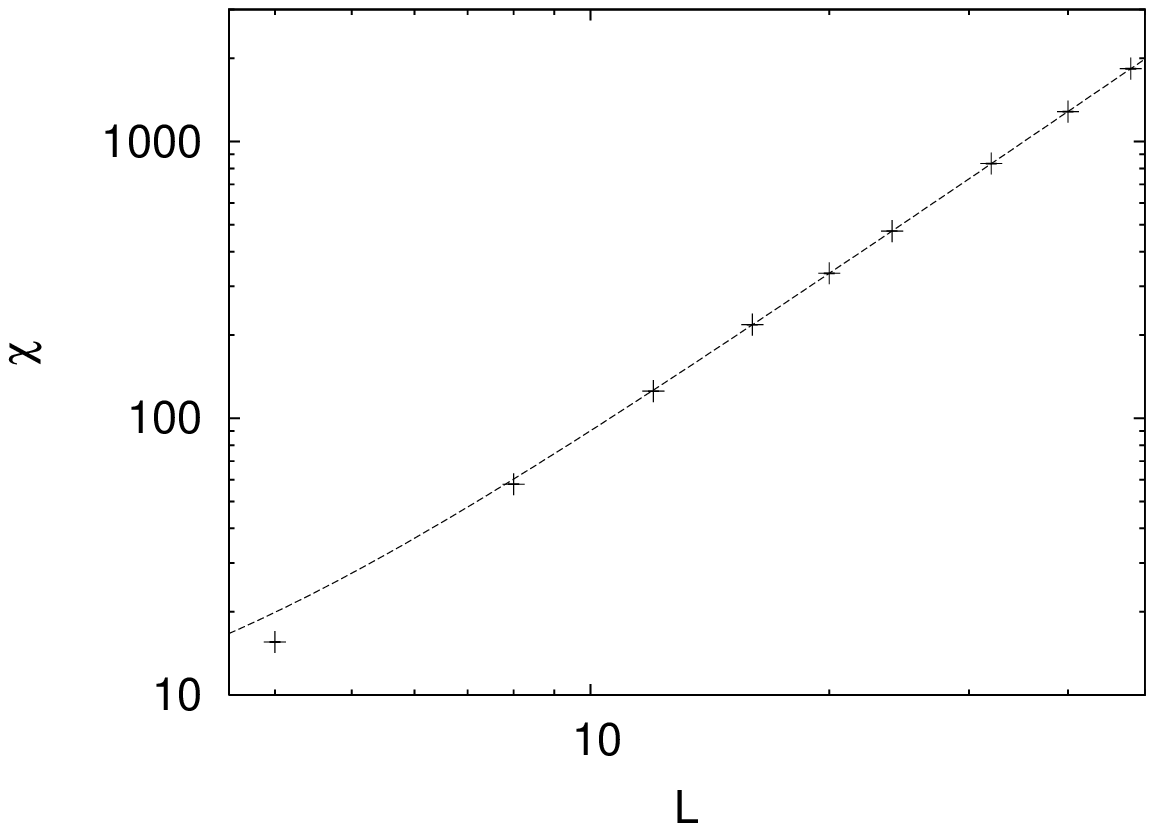,angle=0,height=6.cm,width=7cm}}}
\caption{\label{fig_gambynu}
Log-log plot of the FSS of the susceptibility for $\kappa = 1$ and 
$\sigma = 1.5$ at
$\beta = 0.780\,08 \approx \beta_c$. The
line shows the three-parameter fit $a+b L^{\gamma/\nu}$, yielding for
$L \ge 16$ the estimate $\gamma/\nu=1.962(12)$.
}
\end{figure}

In order to characterize the transition more quantitatively we estimated 
the interface tension\cite{1st_order}, 
\begin{equation}
F_L^s=\frac{1}{2 L^{d-1}}\ln{\frac{P_L^{\rm max}}{P_L^{\rm min}}},
\end{equation}
where $P_L^{\rm max}$ is the value of the two peaks and 
$P_L^{\rm min}$ denotes the minimum in between. Here we have assumed
that for each lattice size the temperature was chosen such that the two 
peaks are of equal height which can be achieved by histogram reweighting. The 
thus defined temperatures approach the infinite-volume transition temperature 
as $1/L^d$, and for the final estimate of $F^s=\lim_{L\rightarrow\infty}F_L^s$,
we performed a fit according to\cite{berg1}
\begin{equation}  
F_L^s = F^s+\frac{a}{L^{d-1}}+\frac{b \ln(L)}{L^{d-1}}.
\label{eq:inter_fss}
\end{equation}
As is shown in Fig.~\ref{fig:inter}, the finite-lattice estimates
$F_L^s$ are clearly nonzero. The infinite-volume extrapolation 
(\ref{eq:inter_fss}) tends to increase with system size and yields a 
comparably large interface tension of $F^s=0.271(5)$.

Let us now turn to the second generic case, $\kappa=1$, where the model 
definitely exhibits for $\sigma = 1.5$ a second-order phase transition around 
$\beta \equiv 1/T \approx 0.8$. To confirm the expected critical exponents of 
the O(2) or XY 
model universality class, we simulated here close to criticality somewhat 
larger lattices of size $L=4,8,12,16,20,24,32,40$, and $48$
and performed a standard FSS analysis. From short runs we 
first estimated the location of the phase transition to be at 
$\beta_0 = 0.7795 \approx \beta_c$. In the long
runs at $\beta_0$ we recorded the time series of the energy density $e=E/N$,
the magnetization $\vec{\mu}$, the mean modulus $\overline{|\psi|}$, and
the mean-square amplitude\cite{remark} $|\psi|^2$, as well as 
the helicity modulus $\Gamma_\mu$ and
the vorticity $v$.
After an initial equilibration 
time we took about $1\,000\,000$ measurements for each lattice size. Applying 
the reweighting technique we first determined the maxima of the susceptibility,
$\chi\prime=N (\langle \vec{\mu}^2\rangle -\langle |\vec{\mu}| \rangle^2)$, 
of $d\langle |\vec{\mu}| \rangle /d\beta$, and of the
logarithmic derivatives $d$ln$\langle |\vec{\mu}| \rangle /d\beta$ and 
$d$ln$\langle \vec{\mu}^2 \rangle/d\beta$.
The locations of these maxima provide us with four sequences of 
pseudo-transition points $\beta_{\rm max}(L)$ for which the scaling 
variable $x=(\beta_{\rm max}(L) - \beta_c) L^{1/ \nu}$ should be constant. 
Using this fact we then have several possibilities to extract the critical 
exponent $\nu$ from (linear) least-squares fits of the FSS ansatz
$dU_L/d\beta \cong  L^{1/ \nu} f_0(x)$ or 
$d$ln$\langle |\vec{\mu}|^p\rangle /d\beta \cong  L^{1/\nu} f_p(x)$ to the 
data at
the various $\beta_{\rm max}(L)$ sequences. The quality of our data and the
fits starting at $L_{\rm min} = 8$, with goodness-of-fit parameters 
$Q=0.85 - 0.90$, can be inspected in Fig.~\ref{fig_nu}. All resulting 
exponent estimates and consequently also their weighted average, 
\begin{equation}
1/\nu=1.493(7), \qquad \nu = 0.670(3),
\end{equation}
are in perfect agreement with recent high-precision Monte Carlo estimates 
for the XY model universality class.\cite{hasen,campostrini} Note that 
hyperscaling implies 
$\alpha = 2 - 3 \nu = -0.010(9)$, which also favorably compares with recent
spacelab experiments on the lambda transition in liquid helium.\cite{g0_helium}

Assuming thus $1/\nu=1.493$ we can improve our estimate for $\beta_c$ from 
linear least-squares fits to the scaling behavior of the various 
$\beta_{\rm max}$ sequences.  The combined estimate from the four sequences is 
$\beta_c = 0.780\,08(4)$.
To extract the critical exponent ratio $\gamma/\nu$ we can now use the scaling 
relation for the susceptibility 
$\chi=N \langle \vec{\mu}^2\rangle \simeq a + b L^{\gamma/\nu}$ at $\beta_c$. 
For $L\ge16$ we obtain from a FSS fit with $Q = 0.70$ the estimate of 
\begin{equation}
\gamma/\nu=1.962(12)[9],
\end{equation}
where we also take into account the uncertainty in our estimate of 
$\beta_c$; this error is estimated by repeating the fit at
$\beta_c\pm\Delta \beta_c$ and indicated by the number in square brackets. 
Here we find a slight dependence of this value 
on the lower bound of the fit range $[L_{\rm min},48]$, i.e., one would have 
to include larger lattices for a high-precision estimate of the critical 
exponent ratio $\gamma/\nu$, but this was not our objective here. Still,
these results are in good agreement with recent high-precision estimates 
in the literature \cite{hasen,campostrini} and clearly confirm the expected second-order
nature of the phase transition in the standard complex $|\psi|^4$ model, 
governed by XY model critical exponents.

A similar set of simulations at $\sigma = 0.25$ for lattice sizes 
$L=4,8,12,14,16,20,24,28,32$, and $40$
gave the
exponent estimates $1/\nu = 1.498(9)$, $\nu = 0.668(4)$ and 
$\gamma/\nu = 1.918(71)[8]$ (at $\beta_c = 0.9284(4)$), 
which are less accurate but again compatible with the XY model universality
class. At any rate these results definitely rule out the possibility of a 
first-order phase transition 
in the standard model
at small $\sigma$-values. When going
to even smaller $\sigma$-values, the FSS analysis is more and more severely 
hampered by 
the vicinity of the Gaussian fixed point which induces strong crossover scaling 
effects. Since consequently very large system sizes would be required to see 
the true, asymptotic (XY model like) critical behavior we have not further
pursued our attempts in this direction. Here we only add the remark that
for $\sigma = 0.01$ the energy and magnetization distributions exhibit a clear
single-peak structure for all considered lattice sizes up to $L=20$, showing
that in the standard model with $\kappa = 1$ a phase-fluctuation induced 
first-order phase transition is very unlikely even for very small $\sigma$ 
values. 

We also checked the critical behavior along the line of second-order
transitions for $\kappa = 0$. Specifically, at $\sigma = 5$, i.e., sufficiently
far away from the crossover to first-order transitions at
$\sigma_t \approx 2.5$,
we obtained from FSS fits to data for lattices of size 
$L=4,8,12,16,20,24,28,32$, and $40$ the exponent estimates
$1/\nu=1.489(7)$, $\nu=0.671(3)$ and
$\gamma/\nu=1.913(82)[13]$ (at $\beta_c = 0.97253(4)$). As expected by
symmetry arguments, also
these results for the second-order regime of the $\kappa = 0$ variant of the
model are in accord with the XY model universality class.

\begin{figure}[tb]
\centerline{\hbox{\psfig{figure=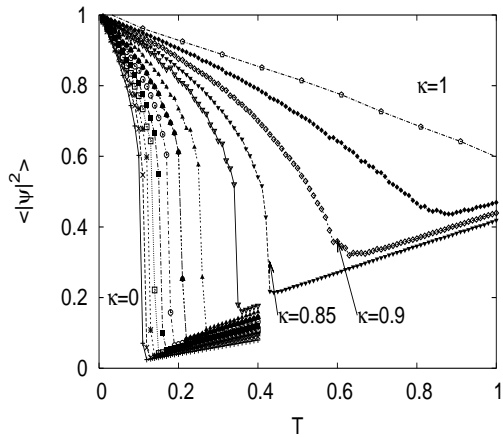,angle=0,height=6.cm,width=7cm}}}
\caption{\label{fig:kappa}
The $\kappa$ dependence of the mean-square amplitude
$\langle |\psi|^2 \rangle$ as a function of temperature on a $15^3$ lattice
for $\sigma=0.25$. 
}
\end{figure}

\begin{figure}[tb]
\centerline{\hbox{\psfig{figure=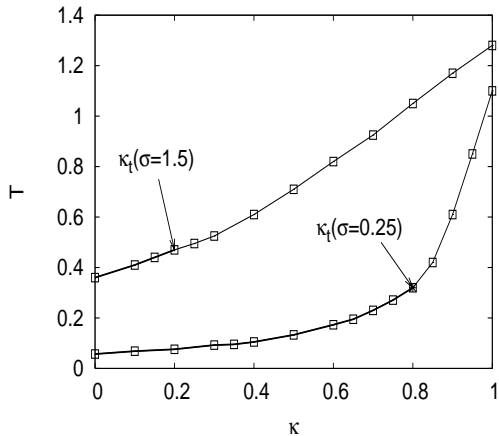,angle=0,height=6.cm,width=7cm}}}
\caption{\label{fig:kappa_T_diag}
Phase diagram in the $\kappa$-$T$-plane of the three-dimensional generalized
complex Ginzburg-Landau model for $\sigma=0.25$ and $\sigma = 1.5$. The 
transitions
along the thick line for $\kappa<\kappa_t$ are of first order, and the 
transitions for $\kappa>\kappa_t$ are of second order. The points labeled
$\kappa_t$ at the intersection of these two regimes are tricritical points.
}
\end{figure}

In a second set of simulations we explored the two-dimensional 
$\sigma$-$\kappa$ parameter space of the generalized Ginzburg-Landau model 
in the orthogonal direction by performing simulations at fixed $\sigma$ values
and $\kappa$ varying from $\kappa = 0$ to $1$. For most $\sigma$-values
we concentrated on the crossover region between first- and second-order
transitions when varying $\kappa$. For two selected values, $\sigma = 0.25$
and $\sigma = 1.5$, we studied the $\kappa$ dependence more systematically 
by simulating all values from $\kappa = 0$ to $1$ in steps of $0.1$. In 
addition we performed two further runs in the crossover regime at 
$\kappa=0.85$ and $0.95$ for $\sigma= 0.25$ as well as at $\kappa = 0.15$ and
$\kappa = 0.25$ for $\sigma = 1.5$. In Fig.~\ref{fig:kappa} we show the 
resulting mean-square amplitudes for all simulated values of 
$\kappa$ at $\sigma = 0.25$ as a function of the temperature, indicating 
again that for small $\kappa$ the transitions are first-order like while 
for $\kappa$ closer to unity the expected second-order transitions emerge.
From Fig.~\ref{fig:kappa} we read off that for $\sigma = 0.25$ the
crossover between the two types of phase transitions happens around
$\kappa_t(\sigma=0.25) \approx  0.8$, and the analogous analysis for 
$\sigma = 1.5$ yields $\kappa_t(\sigma=1.5) \approx 0.2$. The resulting
transition lines for these two $\sigma$-values are plotted in 
Fig.~\ref{fig:kappa_T_diag},
where the thick lines indicate again first-order phase transitions.

\begin{figure}[htb]
%\centerline{\hbox{\psfig{figure=phase.ps,angle=0,height=6.cm,width=7cm}}}
\centerline{\hbox{\psfig{figure=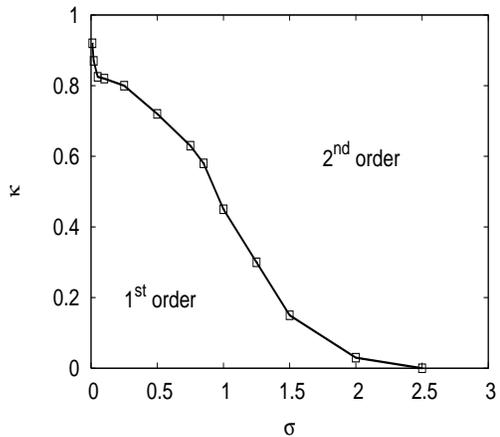,angle=0,height=6.0cm,width=7.0cm}}}
\caption{\label{fig:sigma_kappa_diag}
Phase structure in the $\sigma$-$\kappa$-plane of the generalized complex
Ginzburg-Landau 
model in three dimensions,
separating regions with first- and second-order phase transitions, respectively,
when the temperature is varied.
All continuous transitions fall into
the universality class of the XY model which is approached for all 
$\kappa$-values in the limit $\sigma \longrightarrow \infty$.
}
\end{figure}

Finally, by combining all numerical evidences collected so far with additional
data not described here in detail, we find 
the phase structure in the $\sigma$-$\kappa$-plane depicted in 
Fig.~\ref{fig:sigma_kappa_diag}. All points in the lower left corner for small 
$\sigma$ and small $\kappa$ exhibit temperature driven first-order phase 
transition when the temperature is varied, while all points in the upper right
corner display a continuous transition of the XY model type. This means in 
particular that for the standard model parameterized by $\kappa=1$ this is 
always true. Quantitatively the XY model is reached for all $\kappa$-values 
in the limiting case $\sigma \longrightarrow \infty$.


\subsection{Two dimensions}

% and \ref{2d_fig_v}.

\begin{figure*}[t]
\centerline{\hbox{\psfig{figure=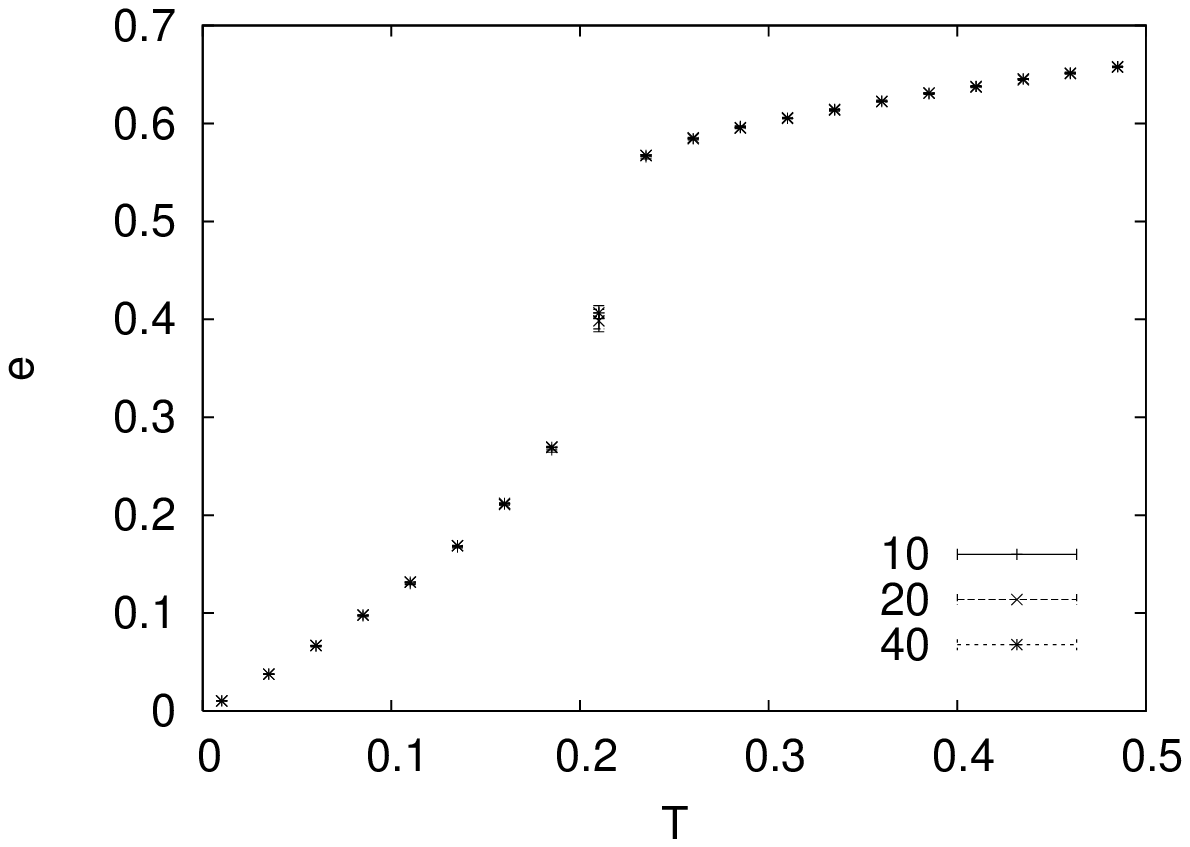,angle=0,height=6.cm,width=7cm}
                  \psfig{figure=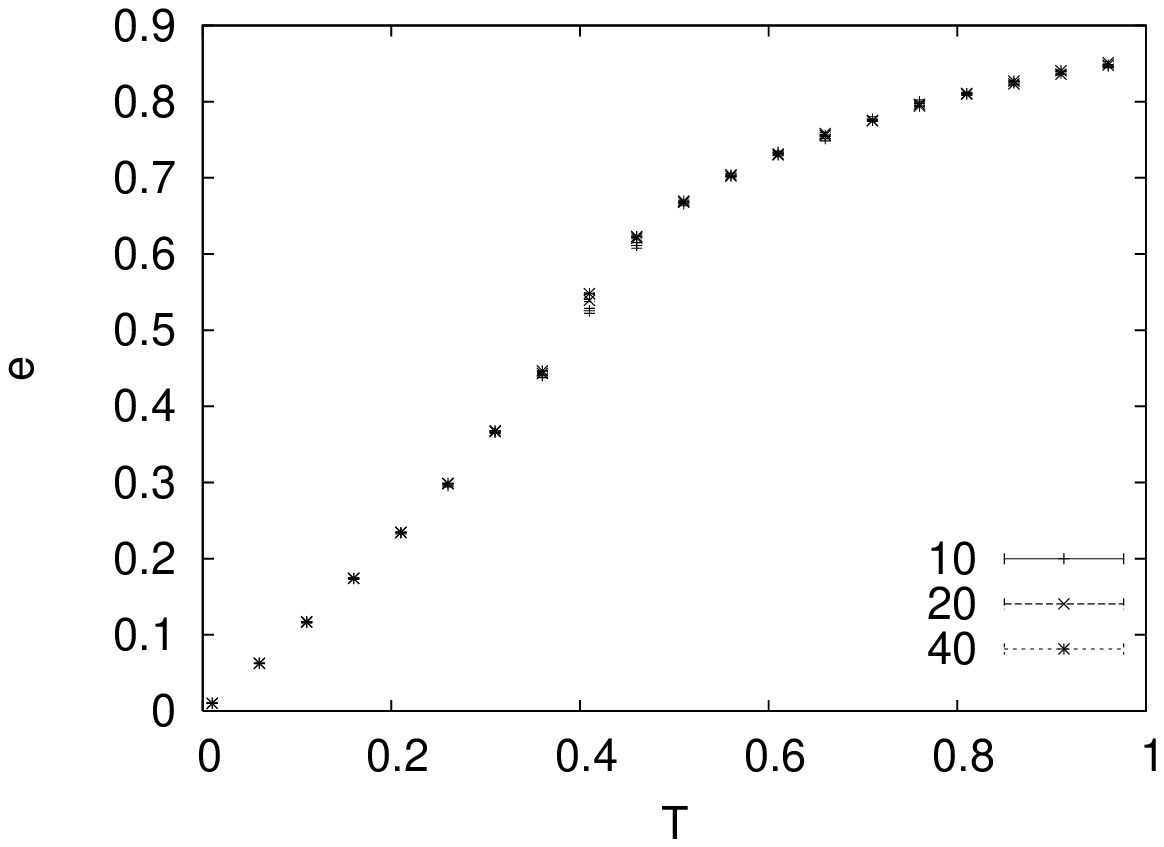,angle=0,height=6.cm,width=7cm}}}
\centerline{\hbox{\psfig{figure=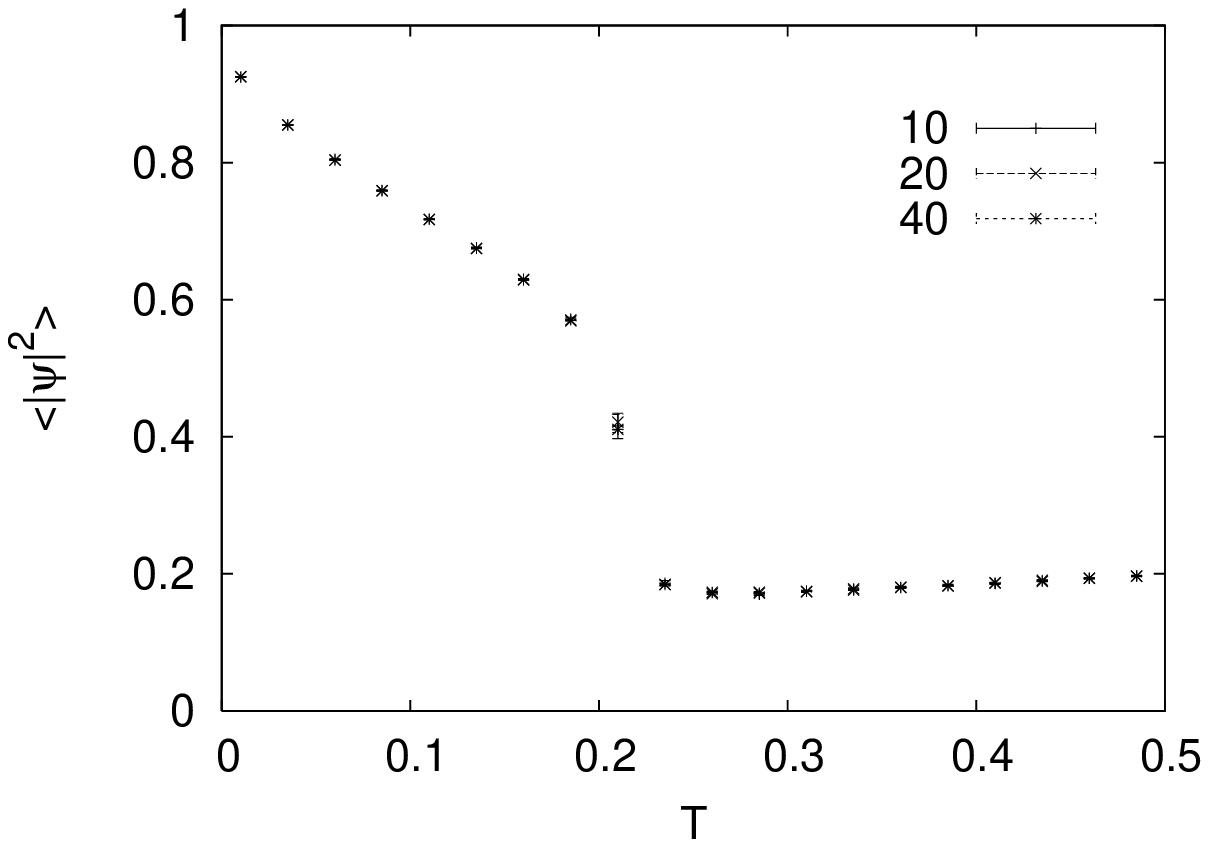,angle=0,height=6.cm,width=7cm}
                  \psfig{figure=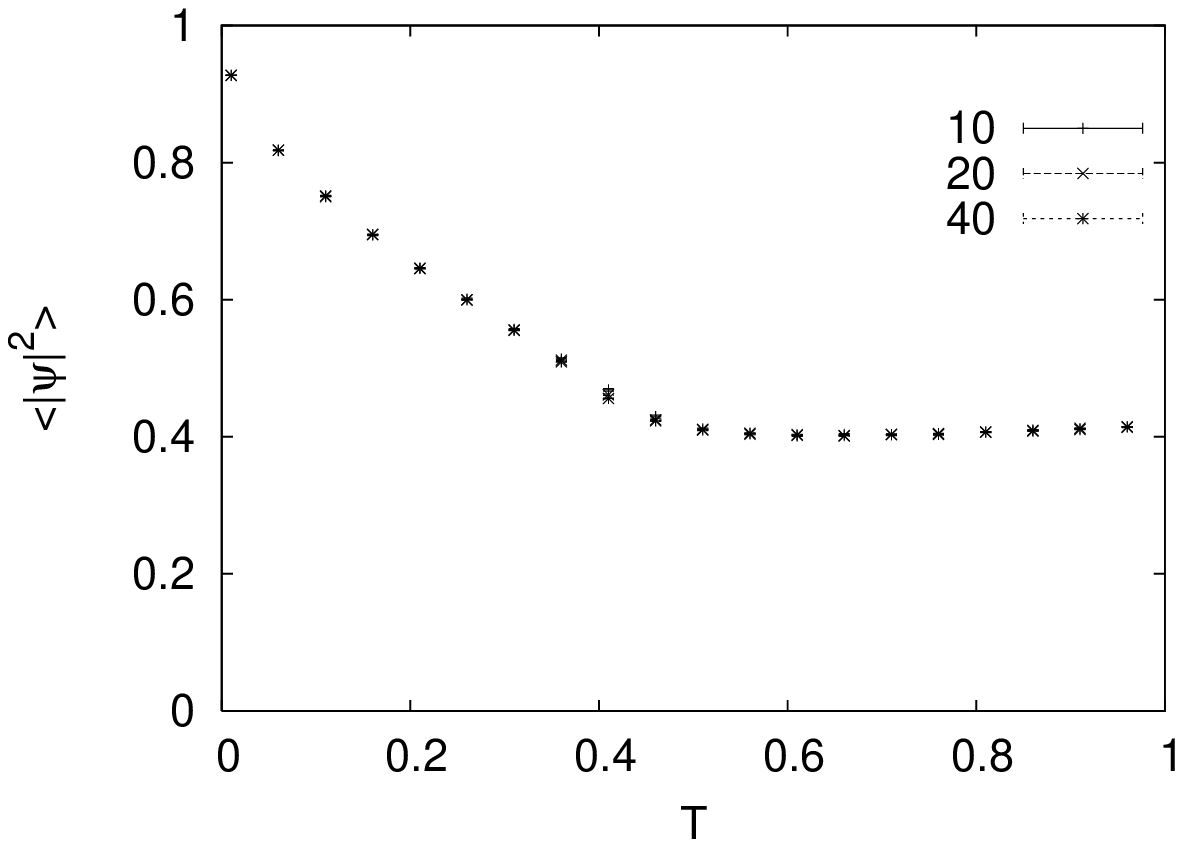,angle=0,height=6.cm,width=7cm}}}
\centerline{\hbox{\psfig{figure=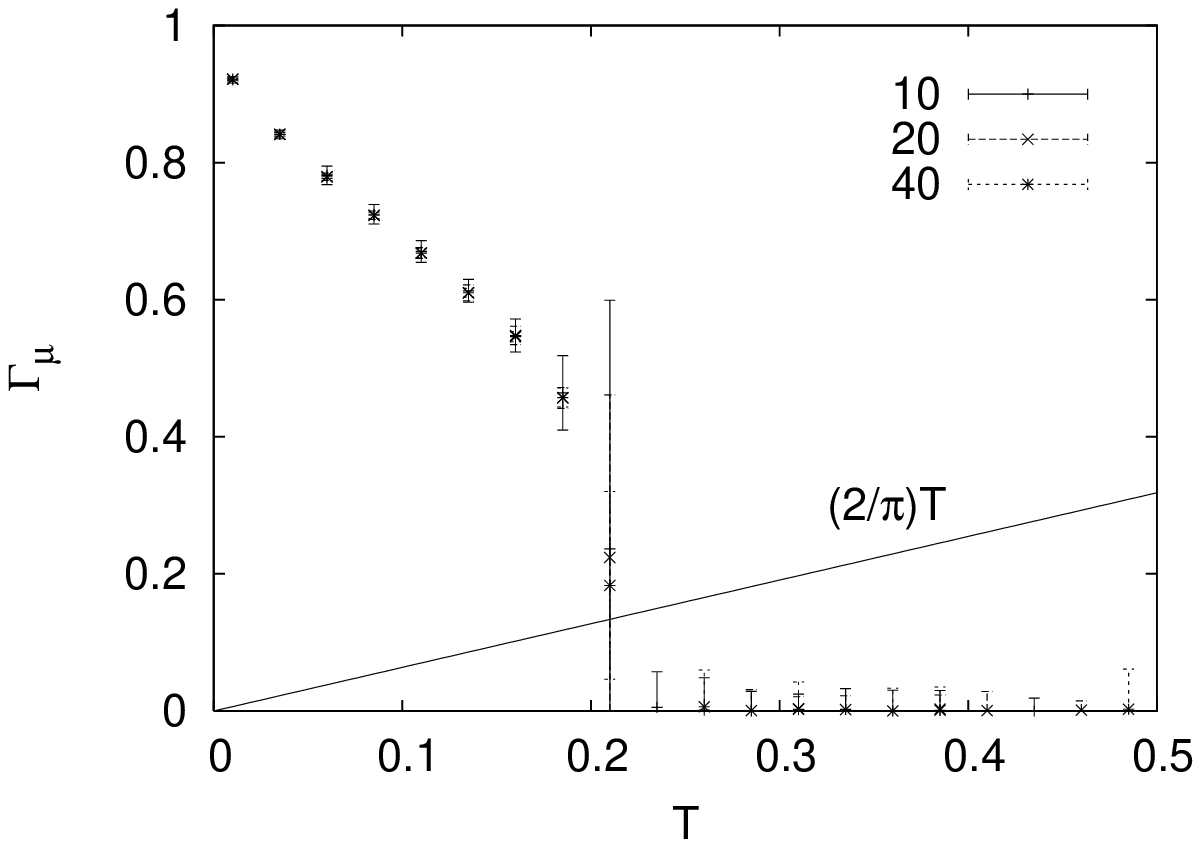,angle=0,height=6.cm,width=7cm}
                  \psfig{figure=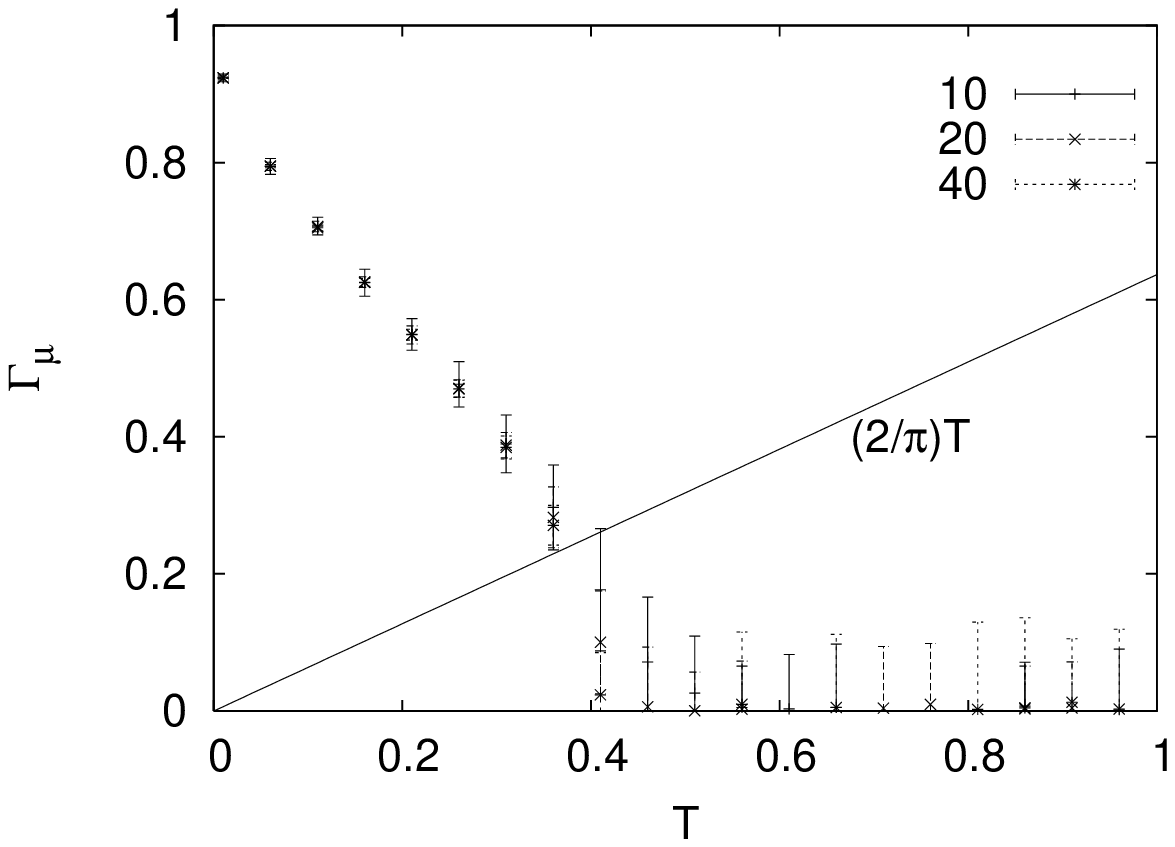,angle=0,height=6.cm,width=7cm}}}
\centerline{\hbox{\psfig{figure=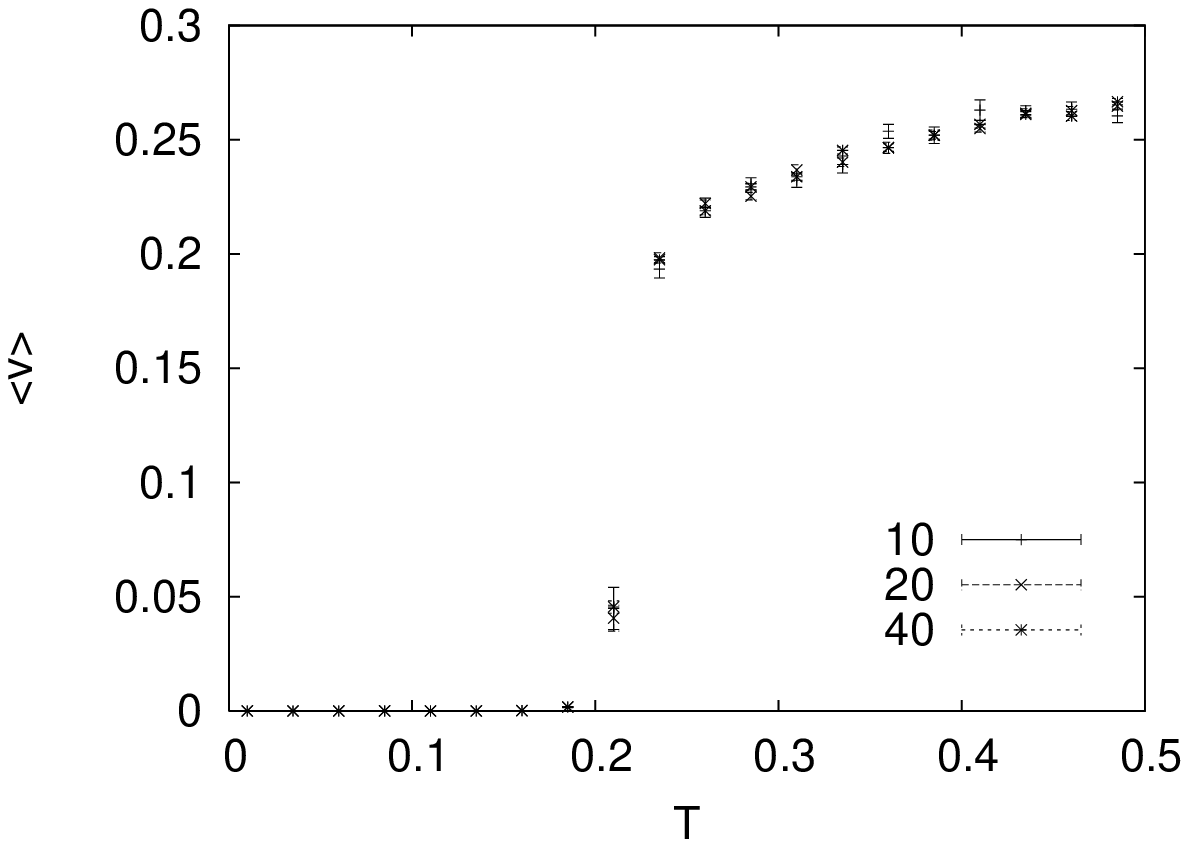,angle=0,height=6.cm,width=7cm}
                  \psfig{figure=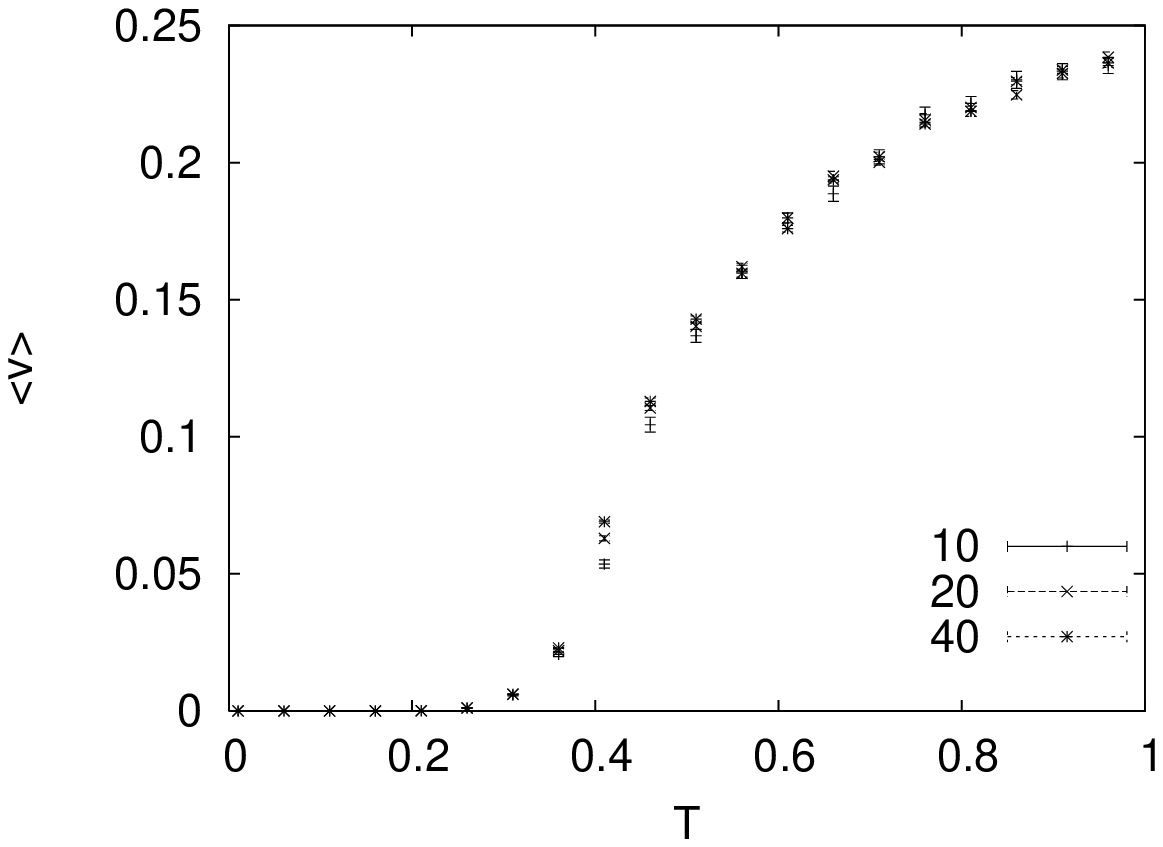,angle=0,height=6.cm,width=7cm}}}
\caption{\label{2d_fig_e}
Energy density $e$,
mean-square amplitude $\langle|\psi|^2\rangle$,
helicity modulus $\Gamma_\mu$ and vortex density $v$ of the two-dimensional 
model on $10^2, 20^2$ and $40^2$ square lattices for $\sigma = 1$ and 
$\kappa=0$ (left) respectively $\kappa=1$ (right).
The straight line in the $\Gamma_{\mu}$ plots
indicate the universal KT jump $\Gamma_{\mu} = (2/\pi) T$ at
$T = T_c$, which clearly is only compatible with the data for the standard model
with $\kappa = 1$.
}
\end{figure*}

We conclude the paper with a few very brief remarks on the two-dimensional
generalized model where the Kosterlitz-Thouless nature of the standard 
XY model transition
would require more care for a precise study. Here we only report results
of some runs at $\sigma=1$ for $10^2$, $20^2$, and $40^2$ square lattices.
As the main result, we find that the standard observables $e$,
$\langle |\psi|^2 \rangle$, $\Gamma$, and $v$ 
exhibit qualitatively the same pattern as in three dimensions. This is
demonstrated in Fig.~\ref{2d_fig_e} where again the two cases $\kappa = 0$
and $\kappa = 1$ are compared. 
For $\kappa = 0$, the data are indicative of a first-order transition
around $T \approx 0.2$, while the behavior of the standard model with 
$\kappa = 1$ is consistent with the expected Kosterlitz-Thouless transition
around $T \approx 0.4$. Note in particular that (only) the data for $\kappa = 1$
are compatible with the expected universal jump of the helicity modulus at $T_c$,
$\Gamma_\nu = (2/\pi) T$, 
indicated by the straight line in the corresponding plots.
A careful investigation of the first-order transitions in the generalized
model with $\kappa = 0$ will be reported elsewhere.

%%%%%%%%%%%%%%%%%%%%%%%%%%%%%%%%%%%%%%%%%%%%%%%%%%%%%%%%%%%%%%%%%%%%%%%%%%
\section{Summary} \label{summary} 
%%%%%%%%%%%%%%%%%%%%%%%%%%%%%%%%%%%%%%%%%%%%%%%%%%%%%%%%%%%%%%%%%%%%%%%%%%

The possibility of a phase-fluctuation induced first-order
phase transition in the standard three-dimensional Ginzburg-Landau model as
suggested by approximate variational calculations \cite{beck1}
cannot be confirmed by our numerical simulations down to very small
values of the parameter $\sigma$. Our results suggest, 
however, that a generalized Ginzburg-Landau model can be tuned to
undergo first-order transitions by a mechanism similar to that
discussed in Ref.~\onlinecite{JK1} when varying the parameter $\kappa$ 
of an additional $\sum \ln R_n$ term in the generalized 
Hamiltonian (\ref{eq:H_gen}). As in Ref.~\onlinecite{JK1} this can be 
understood by a duality argument. For $0 \le \kappa < 1$ the extra term 
reduces the ratio of core energies of vortex lines of vorticity two 
versus those of vorticity one, and this leads to the same type of 
transition as observed in defect melting of crystals.  

The phase transitions of the standard model 
% for all $\sigma$ values 
as well as the continuous transitions of the generalized model are confirmed 
to be governed by the critical exponents of the XY model or O(2) universality
class, as expected by general symmetry arguments. For the generalized model 
it would be interesting to analyze in more detail the tricritical points 
separating the regions with first- and second-order phase transitions. Such
a study, however, is quite a challenging project and hence left 
for the future.

Exploratory simulations of the two-dimensional case, where the
standard model exhibits Kosterlitz-Thouless transitions, indicate
that a similar mechanism can drive the transition of the generalized model
to first order also there.

%%%%%%%%%%%%%%%%%%%%%%%%%%%%%%%%%%%%%%%%%%%%%%%%%%%%%%%%%%%%%%%%%%%%%%%%%%
\section{Acknowledgments}
%%%%%%%%%%%%%%%%%%%%%%%%%%%%%%%%%%%%%%%%%%%%%%%%%%%%%%%%%%%%%%%%%%%%%%%%%%

E.B.\ thanks the EU network HPRN-CT-1999-00161 EUROGRID -- ``Geometry and 
Disorder: from membranes to quantum gravity'' for a postdoctoral grant. 
Partial support by the German-Israel-Foundation (GIF) under contract 
No.\ I-653-181.14/1999 is also gratefully acknowledged.


\begin{thebibliography}{99}

\bibitem{ZinnJ} J. Zinn-Justin, {\em Quantum Field Theory and Critical
Phenomena\/}, 3rd ed. (Clarendon Press, Oxford, 1996).
\bibitem{KlSch01} H. Kleinert and V. Schulte-Frohlinde, {\em Critical
Properties of $\Phi^4$-Theories\/} (World Scientific, Singapore, 2001).
\bibitem{Kl00} H. Kleinert, Phys. Rev. Lett. {\bf 84}, 286 (2000).
\bibitem{beck1} P. Curty and H. Beck, Phys. Rev. Lett. {\bf 85}, 796 (2000).
\bibitem{beck2} P. Curty and H. Beck, cond-mat/0010084.
\bibitem{fort2d_0} H. Fort, hep-th/0010070.
\bibitem{fort2da} G. Alvarez and H. Fort, Phys. Rev. {\bf B63}, 132504 (2001).
% cond-mat/0006341.
\bibitem{fort2db} G. Alvarez and H. Fort, Phys. Lett. {\bf A282}, 399 (2001). 
% cond-mat/0010137.
\bibitem{fort3d} G. Alvarez and H. Fort, Phys. Rev. {\bf B64}, 092506 (2001).
% cond-mat/0010119.
\bibitem{footnote1} The method of Ref.~\onlinecite{beck1} uses XY model data 
as input.
\bibitem{Kl89} H. Kleinert, {\em Gauge Fields in Condensed Matter\/},
Vol.~I: {\em Superflow and Vortex Lines\/} (World Scientific, Singapore, 1989).
\bibitem{LoQuSh01} V.M. Loktev, R.M. Quick, and S.G. Sharapov, Phys. Rep.
{\bf 349}, 1 (2001).
\bibitem{ebwj_prl} E. Bittner and W. Janke, Phys. Rev. Lett. {\bf 89}, 130201 (2002).
\bibitem{footnote2}
There is another difference between our update and the
update described in the papers \cite{beck2,fort3d}, namely that we do
{\em not\/} restrict the modulus of the field to a finite interval. This
can cause further systematic deviations but we explicitly checked that this
is unimportant for the main point here.
\bibitem{JK1} W. Janke and H. Kleinert,
%    {\em First Order Phase Transition in 3D XY Model with Mixed Action,\/},
   Nucl. Phys. {\bf B270}, 399 (1986).
\bibitem{MiWa87} P. Minnhagen and M. Wallin, Phys. Rev. {\bf B36}, 5620 (1987);
G.-M. Zhang, H. Chen, and X. Wu, Phys. Rev. {\bf B48}, 12304 (1993);
D.Yu. Irz, V.N. Ryzhov, and E.E. Tareyeva, Phys. Rev. {\bf B54}, 3051 (1996).
\bibitem{Kl89II} H. Kleinert, {\em Gauge Fields in Condensed Matter\/},
Vol.~II: {\em Stresses and Defects\/} (World Scientific, Singapore, 1989).
\bibitem{WJ_defect} W. Janke, Int. J. Theor. Phys. {\bf 29}, 1251 (1990).
\bibitem{Metro} N. Metropolis, A.W. Rosenbluth, M.N. Rosenbluth,
A.H. Teller, and E. Teller,
% \textit{ Equation of State Calculations by Fast Computing Machines},
J. Chem. Phys. {\bf 21}, 1087 (1953).
% --1092.
\bibitem{WJ_review} W. Janke, Mathematics and Computers in Simulations
{\bf 47},  329 (1998); and in:
{\em Computational Physics: Selected Methods -- Simple Exercises --
Serious Applications\/}, eds. K.H. Hoffmann and M. Schreiber (Springer,
Berlin, 1996); p.~10.
\bibitem{wolff} U. Wolff, Phys. Rev. Lett. {\bf 62}, 361 (1989); Nucl.
Phys. {\bf B322}, 759 (1989).
\bibitem{hasen} M. Hasenbusch and T. T\"or\"ok,
%{\em High precision Monte Carlo study of the 3D XY-universality class},
J. Phys. {\bf A32}, 6361 (1999).
\bibitem{WJ_XY} W. Janke, Phys. Lett. {\bf A148}, 306 (1990).
% cond-mat/9904408.
\bibitem{berg}
B.A. Berg and T. Neuhaus, Phys. Lett. {\bf B267}, 249
 (1991); Phys. Rev. Lett. {\bf 68}, 9 (1992).
\bibitem{Jack} B. Efron, {\em The Jackknife, the Bootstrap and Other
Resampling Plans} (Society for Industrial and Applied Mathematics [SIAM],
Philadelphia, 1982).
\bibitem{1st_order}
W. Janke, {\em First-Order Phase Transitions\/},
in: {\em Computer Simulations of Surfaces and Interfaces\/},
NATO Science Series, II. Mathematics, Physics and Chemistry --
Vol.~{\bf 114}, Proceedings of the NATO Advanced Study Institute, 
Albena, Bulgaria, 9--20 September 2002, edited by B. D\"unweg,
D.P. Landau, and A.I. Milchev (Kluwer, Dordrecht, 2003), pp.~111--135.
\bibitem{berg1}
U. Hansmann, B.A. Berg, and T. Neuhaus, Int. J. Mod. Phys. {\bf C3}, 1155 (1992).
\bibitem{remark}
Recall that $\overline{|\psi|} \equiv \sum_{n=1}^N |\psi_n|/N$ and
$|\psi|^2 \equiv \sum_{n=1}^N |\psi_n|^2/N$, such that
$\overline{|\psi|}^2 \ne |\psi|^2$.
\bibitem{campostrini}
M. Campostrini, M. Hasenbusch, A. Pelissetto, P. Rossi, and E. Vicari,
%{\em Critical behavior of the three-dimensional XY universality class}, 
Phys. Rev. {\bf B63}, 214503 (2001).
\bibitem{g0_helium} J.A. Lipa, J.A. Nissen, D.A. Stricker, D.R. Swanson, and
T.C.P. Chui, 
% Specific heat of liquid helium in zero gravity very near the lambda point
Phys. Rev. {\bf B68}, 174518 (2003).
% [25 pages]

\end{thebibliography}
\end{document}